\documentclass[fleqn,usenatbib]{mnras}

\usepackage{newtxtext,newtxmath}

\usepackage[T1]{fontenc}

\DeclareRobustCommand{\VAN}[3]{#2}
\let\VANthebibliography\thebibliography
\def\thebibliography{\DeclareRobustCommand{\VAN}[3]{##3}\VANthebibliography}

\usepackage{ulem}

\usepackage{graphicx}	
\usepackage{amsmath}	

\title[Chemical enrichment in cosmological simulations]{A subgrid model for chemical enrichment in cosmological simulations}
\author[C. A. Correa et al.]{Camila A. Correa$^{1}$\thanks{E-mail: camila@camilacorrea.com}, Joop Schaye$^{1}$, Matthieu Schaller$^{2,1}$, James W. Trayford$^{3}$, Evgenii Chaikin$^{1}$,
\newauthor Alejandro Benitez-Llambay$^{4}$, Carlos S. Frenk$^{5}$, Sylvia Ploeckinger$^{6}$, and Alexander J. Richings$^{7,8}$ \\
$^{1}$ Leiden Observatory, Leiden University, PO Box 9513, 2300 RA Leiden, the Netherlands \\
$^{2}$ Lorentz Institute for Theoretical Physics, Leiden University, PO Box 9506, 2300 RA Leiden, the Netherlands \\
$^{3}$ Institute of Cosmology and Gravitation, University of Portsmouth, Dennis Sciama Building, Burnaby Road, Portsmouth PO1 3FX, UK \\
$^{4}$ Dipartimento di Fisica G. Occhialini, Università degli Studi di Milano Bicocca, Piazza della Scienza, 3 I-20126 Milano MI, Italy \\
$^{5}$ Institute for Computational Cosmology, Department of Physics, University of Durham, South Road, Durham, DH1 3LE, UK \\
$^{6}$ Department of Astrophysics, University of Vienna, T\"urkenschanzstrasse 17, 1180 Vienna, Austria \\
$^{7}$ Centre for Data Science, Artificial Intelligence and Modelling, University of Hull, Cottingham Road, Hull, HU6 7RX, UK \\
$^{8}$ E. A. Milne Centre for Astrophysics, University of Hull, Cottingham Road, Hull, HU6 7RX, UK
}

\date{Accepted XXX. Received YYY; in original form ZZZ}

\pubyear{2025}

\begin{document}

\label{firstpage}
\pagerange{\pageref{firstpage}--\pageref{lastpage}}
\maketitle

\begin{abstract}
We present the modules for stellar nucleosynthesis, stellar mass loss, and turbulent diffusion of the new COLIBRE subgrid model for cosmological hydrodynamical simulations of galaxy formation. COLIBRE models the thermal evolution of the multi-phase interstellar medium, dust grains, star formation, and stellar and AGN feedback. This work focuses on the model for chemical enrichment. We track the evolution of 12 chemical elements produced by a broad range of nucleosynthetic channels, including core-collapse supernovae and stellar winds, Type Ia supernovae, and asymptotic giant branch (AGB) stars. Enrichment from $s$- and $r$-process elements is modelled via contributions from AGB stars, neutron star mergers, common envelope supernovae, and collapsars. We present an updated compilation of stellar yields taken from the literature, which we release alongside this work\footnotemark. Small-scale element mixing is implemented through a turbulent diffusion process. While diffusion has only a minimal impact on basic integrated galaxy properties, it does reduce the slope of the gas-phase metallicity-mass relation compared with simulations that do not include it. The distribution of element ratios of individual stellar particles is sensitive to diffusion, but only at low metallicities ($Z \lesssim 10^{-1}\,\rm{Z}_\odot$). The model is tested using redshift $z=0$ results from a set of cosmological simulations, mostly of (25 Mpc)$^3$ volumes, demonstrating generally good agreement with Milky Way stellar abundance trends from the APOGEE survey. The model also reproduces the alpha-element enhancement relations observed in galaxies from SDSS, ATLAS-3D, and the Local Group.
\end{abstract}

\footnotetext{\url{https://github.com/correac/COLIBRE_yield_tables.git}}


\section{Introduction}

The chemical composition of gas and stars within galaxies provides valuable insights into the physical processes involved in galaxy formation. These processes include the stochastic nature of star formation (e.g., \citealt{Ellison08,Mannucci10,Curti20}), the recycling of gas through inflows and metal-rich outflows (e.g. \citealt{Tremonti04, Erb06, Oppenheimer06,Mitchell20,vanloon21}), the dependence of nucleosynthetic yields on mass and metallicity (e.g. \citealt{Romano10,Nomoto13}), the efficiency of gas mixing in the interstellar medium (ISM) (e.g. \citealt{Escala18,Krumholz18,Emerick20}), and the dependence of gas cooling on elemental abundances (e.g. \citealt{Sutherland93, Gnat07, Wiersma09b}). 

The chemical composition of galaxies reflects the total enrichment of gas over their entire star formation history, and hence acts as a fossil record. This area of astrophysics is currently flourishing. Large spectroscopic surveys such as LAMOST (\citealt{Cui12}), Gaia-ESO (\citealt{Gilmore12,Randich22}), Gaia DR3 (\citealt{RecioBlanco23}), APOGEE (\citealt{Majewski17,Ahumada20,Abdurrouf22}), GALAH (\citealt{DeSilva15,Buder21}), 4MOST (\citealt{Chiappini19}), DESI (\citealt{Zhang24}), WEAVE (\citealt{Iovino23}), Pristine (\citealt{Martin23}), and MOONS (\citealt{Gonzalez20}) are providing extensive observational data for studying stellar chemical abundances. Furthermore, integral field unit spectrographs, such as CALIFA (e.g. \citealt{Sanchez12}), SAMI (e.g. \citealt{Sanchez19}), MaNGA (e.g. \citealt{Belfiore17}), KMOS (e.g. \citealt{Wuyts14, Wuyts16}), MUSE (e.g. \citealt{Sarzi18, Erroz19}), and NIRSpec on JWST (e.g. \citealt{Curti23}) are enabling spatially resolved measurements of metallicity within galaxies.

To keep up with the analysis and exploitation of these high-quality observational datasets, improved models for galactic chemical evolution are needed. Numerous galactic chemo-dynamical models have been developed and refined over the years (e.g. \citealt{Mosconi01,Kobayashi07,Prantzos08,Matteucci12,Dave12,Lilly13,Sharda24}). These models range from analytical approaches (e.g. \citealt{Molla15,Andrews17,Rybizki17,Weinberg17,Cote17,Spitoni17,Ritter18,Johnson20,Gjergo23}) to hydrodynamical simulations that incorporate detailed chemical enrichment and galactic dynamics (e.g. \citealt{Kawata03, Tornatore07, Finlator08, Wiersma09, Few12, Buck21, Kobayashi20}).

Particularly relevant to this work are cosmological hydrodynamical simulations of galaxy formation, which have been remarkably successful in producing populations of galaxies with realistic properties (e.g. \citealt{Schaye15, Pillepich19, Dave19}). These simulations can replicate various chemo-dynamical properties of galaxies (e.g. \citealt{Hirschmann16, Grand18, Agertz20, Hough23}). However, they incorporate stellar physics components that are rather uncertain (e.g. \citealt{Cote16}). 

One example of such an uncertain ingredient is the stellar nucleosynthesis yields from core-collapse supernovae (CCSNe) or asymptotic giant branch (AGB) stars (e.g., \citealt{Wiersma09, Romano10, Nomoto13, Pignatari16}). For instance, \citet{Philcox18} concluded that the stellar yields from \citet{Prantzos18} for AGB stars, and \citet{Chieffi04} for CCSNe best reproduce proto-solar abundances. In contrast, \citet{Buck21} showed that within the framework of the NIHAO feedback model (\citealt{Wang15}), the parameter combination that best  reproduces observed stellar abundances consists of the yield tables from \citet{Karakas16} for AGB stars, \citet{Chieffi04} for CCSNe, and \citet{Seitenzahl13} for supernova Ia (SNIa) along with a power-law delay time distribution (DTD) function following \citet{Maoz12}, where the DTD function describes the distribution of stellar ages when the Type Ia supernovae explode.

The SNIa delay times are also uncertain. Recently, \citet{Cavichia24} explored 15 DTD functions along with 12 tables of elemental yields produced by different SNIa explosion mechanisms. They found that a DTD with a prompt component can simultaneously reproduce the whole set of compiled data from the Milky Way related to the age-metallicity relation, the iron-to-hydrogen ratio [Fe/H]\footnote{Throughout this work we use the notation [X/Y]$\equiv \log_{10}(n_{\rm{X}}/n_{\rm{Y}})-\log_{10}(n_{\rm{X}}/n_{\rm{Y}})_{\odot}$, where $n_{\rm{X}}$ is the number density of species X and solar abundances are taken from \citet{Asplund09}.}, and the relative abundances of $\alpha$-elements ([$\alpha$/Fe]) $-$where $\alpha$-elements include O, Mg, Si$-$ as functions of stellar age or metallicity. Similarly, \citet{Dubay24} implemented various DTD forms in combination with different star formation histories for the Milky Way. They concluded that the DTD alone cannot explain the observed bimodality in the [$\alpha$/Fe] distribution. Their model suggests that an extended DTD with fewer prompt SNIa is needed to reproduce the observed MW [$\alpha$/Fe] ratios.

The stellar initial mass function (IMF) is also a key component of chemo-dynamical models. It influences most observable properties of stellar populations by regulating the relative fractions of low- and high-mass stars within them. However, it is still uncertain whether the IMF is universal and invariant or varies with the local physical conditions (see e.g.\ \citealt{Smith20,Hennebelle24}). 

Another crucial component of chemo-dynamical models is metal mixing in the interstellar medium (ISM) caused by gas turbulence. Numerical codes using smoothed particle hydrodynamics (SPH), like the one in this study, lack inherent numerical mixing and therefore require additional transport terms. While this extra subgrid modeling allows precise control over mixing strength, the challenge lies in understanding how much mixing occurs naturally in the ISM (see e.g. \citealt{Greif09, Shen10, Petit15, Rennehan19, Rennehan21}). Grid codes, on the other hand, may suffer from excessive numerical mixing (e.g. \citealt{Wadsley08, Schmidt15}).

In this study, we aim to explore these uncertain physical components. To this end, we introduce the chemical enrichment model implemented in the new COLIBRE galaxy formation model for cosmological hydrodynamical simulations. COLIBRE includes the modeling of a multi-phase interstellar medium, non-equilibrium gas cooling, a live dust model integrated with the chemistry, and detailed prescriptions for star formation and feedback from star formation and active galactic nuclei (AGN). More information on the features of the COLIBRE model and the calibration strategy is provided in \citet{Schaye25} and \citet{Chaikin25}, respectively.

In this work, we investigate how stellar nucleosynthetic yields and metal mixing affect the evolution of galaxies. A complementary study by Nobels et al. (in prep.) examines how different SNIa DTDs influence galaxy properties and cosmic metal enrichment. The impact of the IMF will be addressed in future work.

This paper is organized as follows. Section \ref{Sec2} described the chemical enrichment model implemented in COLIBRE, with specific details on the nucleosynthesis yields (Section \ref{Sec21}) and the treatment of metal mixing (Section \ref{Metal_diffusion_Sec}). Section \ref{Sec3} summarizes the main aspects of the COLIBRE model, along with the simulation series produced for this work. Section \ref{Sec4} deals with the uncertainties in the model, (re-)normalization of some yields (Section \ref{CCSN_boost_factors_section}) and mixing rate (Section \ref{Sec42}). We present our results in Section \ref{Sec5} and conclude in Section \ref{Sec6}.

\section{Chemical enrichment model}\label{Sec2}

\begin{table*}
\caption{Literature sources for the nucleosynthesis yields $Y_{k,\rm{tbl}}(M_{\rm{tbl}},Z_{\rm{tbl}})$. The first and second columns from the left indicate the enrichment source and the corresponding reference from which the yields were obtained. The middle and right columns show the zero-age main sequence masses and metallicities, expressed as absolute mass fractions.}
\begin{center}
\begin{tabular}{ l|c|c|c }
\hline
Yield table & Type & Masses $M_{\rm{tbl}}$ [M$_{\odot}$] & Metallicities $Z_{\rm{tbl}}$ \\
\hline\hline
\citet{Karakas10} &  AGB  & [$1-6$] & [0.0001, 0.004]  \\
\citet{Fishlock14} &  AGB  & [$1-6$] & [0.001]  \\
\citet{Doherty14} &  AGB  & [$6.5-9$] & [0.004, 0.008, 0.02]  \\
\citet{Karakas16} &  AGB  & [$1-8$] & [0.007, 0.014, 0.03]  \\
\citet{Cinquegrana22} &  AGB  & [$1-8$] & [0.04, 0.05, 0.06, 0.07, 0.08, 0.09, 0.1]  \\
\hline
\citet{Leung18}, W7 & SNIa & [1.37] & [0.0134] \\
\hline
\citet{Nomoto13} & CCSN & [$13-40$] & [0.001, 0.004, 0.008, 0.02, 0.05] \\
\hline
\end{tabular}
\end{center}
\label{Yield_tables}
\end{table*}

In our model, each star particle represents a Simple Stellar Population (SSP) following the IMF of \citet{Chabrier03} within the mass range $0.1-100~\rm{M}_{\odot}$. Because of the limited resolution of the simulations the effects of stochastically sampling the IMF would be small, and we thus assume each stellar particle to contain the average mass fraction of each stellar mass. We assume the IMF to be universal, with the high- and low-mass slopes not changing with mass, metallicity or time. While this is a standard simplifying assumption, the IMF is highly uncertain in more extreme environments and minor variations in the IMF can have significant implications for global galaxy properties (see e.g. \citealt{Barber19a,Barber19b}).  	

Each star particle $i$ is characterized by its time of birth, $t_{{\rm{birth}},i}$, total mass, $m_{i}$, metallicity, $Z_{i}$, and element composition, $X_{i}$. These parameters are used to determine the stellar lifetimes and nucleosynthesis yields. When a star particle is formed, it inherits the total mass and the elemental abundances of its parent gas particle. This also sets the star particle's birth time $t_{{\rm{birth}},i}$. Mass- and metallicity-dependent lifetimes, $\tau(M,Z_{i})$, from \citet{Portinari98} are employed to calculate the lifetimes of the SSPs, with $M$ denoting the zero age main sequence mass of stars from the SSP. At each time step, $\tau(M,Z_{i})$ is utilized to identify stellar masses reaching the end of the main sequence or exploding as CCSNe. The fraction of the initial particle mass reaching the different evolutionary stages, along with the elemental abundances, is employed to compute the mass of each element lost through winds from AGB stars, winds from massive stars and ejecta from SNIa and CCSNe. Stellar evolution models track many elements, however only eleven contribute significantly to the radiative cooling rates: H, He, C, N, O, Ne, Mg, S, Si, Ca and Fe (\citealt{Wiersma09b}). In addition to these eleven elements, we track Sr, Ba and Eu. All elements are traced explicitly, with the exception of Ca and S, which we assume to be proportional to Si. Throughout this work, we define metallicity, $Z$, as the total mass of all elements heavier than He, whether explicitly tracked or not, divided by the total mass of the star.

Six channels contribute to metal enrichment: SNIa, CCSNe, AGB stars, neutron star mergers (NSM), common envelop jets supernovae (CEJSN) and collapsars. For CCSNe and AGB the elemental enrichment and remnant mass vary with stellar metallicity and mass. SNIa, on the other hand, follow a different explosion path and we do not consider a metallicity dependence.

In our modelling, stars in the (zero age main sequence) mass range $1-8~\rm{M}_{\odot}$ enrich the ISM via the AGB phase, whereas stars in the $8-40~\rm{M}_{\odot}$ range become CCSNe. Stars more massive than $40~\rm{M}_{\odot}$ do not enrich the ISM. We assume that these stars become failed supernovae and end their lives as black holes. Low- and intermediate-mass stars in the AGB phase are the main production sites of $s$-process elements, such as strontium and barium (e.g. \citealt{Karakas14}). NSM, CEJSN and collapsars are the astrophysical sites that we consider for the nucleosynthesis of $r$-process elements, such as europium.

Following the methodology of \citet{Wiersma09}, we calculate the total mass ($\Delta m_{i,\rm{total}}$), total metal mass ($\Delta m_{i,\rm{metal}}$) and mass of each tracked element $k$ ($\Delta m_{i,k}$) that are ejected during a single time step ($\Delta t$) by a star particle $i$ as

\begin{equation}\label{deltami}
\Delta m_{i,{\rm{total/metal}}/k}=m_{i}\int_{M_{Z}(\tau_{i}+\Delta t)}^{M_{Z}(\tau_{i})}\Phi(M)m_{{\rm{ej}},{\rm{total/metal}}/k}(M,Z_{i}){\rm{d}}M,
\end{equation}

\noindent where $M_{Z}(t)$ is the inverse of the metallicity-dependent lifetime function from \citet{Portinari98}, $\tau_i$ is the stellar particle's age, and $\Phi(M)$ is the IMF, normalized such that $\int M\Phi(M){\rm{d}}M=1$. The values of $m_{{\rm{ej}},{\rm{total/metal}}/k}$ are taken from nucleosynthesis tables, and they correspond to the total mass, $m_{\rm{ej,total}}$, the total metal mass, $m_{\rm{ej,metal}}$, or the mass of element $k$, $m_{{\rm{ej}},k}$, ejected by a single star of zero age main sequence mass $M$ and metallicity $Z$. 

At each time step, $\Delta m_{i,{\rm{total/metal}}/k}$ is distributed to the neighbouring gas particles $j$ that reside within the SPH kernel with smoothing length $h$ of the star particle $i$. This is done by weighting the metal distribution based on the gas particles densities and spatial distance within the kernel. The total mass assigned to gas particle $j$, $\Delta m_{j,{\rm{total/metal}}/k}$, is calculated as follows,

\begin{equation}\label{deltami2}
\begin{aligned}
& \Delta m_{j,{\rm{total/metal}}/k} = \\
&\quad\quad\quad \Delta m_{i,{\rm{total/metal}}/k}\frac{W(r_{ij},h)}{\rho_{j}}\left[\sum_{\alpha}\frac{W(r_{i\alpha},h)}{\rho_{\alpha}}\right]^{-1},
\end{aligned}
\end{equation}


\noindent where $W(r_{ij},h)$ is the SPH kernel, $r_{ij}$ the distance between star particle $i$ and gas particle $j$, and $\rho_{j}$ is the density of gas particle $j$. In the denominator, the sum is taken over all neighbouring gas particles. Note that \citet{Wiersma09} used the weights $m_j W(r_{ij},h) / \rho_j$. For EAGLE \citet{Schaye15} replaced the current mass of the receiving particle, $m_j$, by the initial gas particle mass in order to avoid biasing the mass transfer towards gas particles with higher metallicities (gas particles can only increase their mass by receiving mass from star particles, which tends to increase their metallicities). For EAGLE this is equivalent to using our weights of $W(r_{ij},h) / \rho_j$ because the initial gas particles mass is constant in EAGLE and hence drops out of eq.~(\ref{deltami2}). For COLIBRE the initial gas masses vary slightly in the initial conditions. Therefore, to avoid a small bias in the direction of mass loss that depends on the initial conditions, we use $W(r_{ij},h) / \rho_j$ as weights.

Note that when we calculate $\Delta m_{i,\rm{metal}}$ in eq.~(\ref{deltami2}), it corresponds to the total metal mass of all elements heavier than helium, regardless of whether we track them individually or not. Further details on the determination of $m_{\rm{ej,total}}$, $m_{\rm{ej,metal}}$, and $m_{{\rm{ej}},k}$, are given in the following subsection, along with details of the nucleosynthesis yields adopted by the model.

\subsection{Nucleosynthesis yields}\label{Sec21}

\subsubsection{AGB yields}\label{AGB_yields_sec}

We assume that stars in the mass range $1-8~\rm{M}_{\odot}$ enrich the ISM via the AGB phase. The final fate of super-AGB stars with initial masses between $9-12~\rm{M}_{\odot}$ is uncertain (\citealt{Doherty17}). It depends on the competition between mass loss, which reduces the envelope mass, and an increase in the core mass through H-He shell burning. Fast mass loss leads to the formation of an O+Ne+Mg white dwarf (e.g. \citealt{Siess07}). Alternatively, $9-12~\rm{M}_{\odot}$ stars can become electron-capture supernovae (\citealt{Wanajo11, Wanajo13}). The metal enrichment due to such supernovae is low (\citealt{Wanajo13}) and is not considered in this work.

We adopt the AGB nucleosynthesis yields from \citet{Karakas10}, \citet{Fishlock14}, \citet{Karakas16}, \citet{Doherty14} and \citet{Cinquegrana22}. Table~\ref{Yield_tables} indicates the mass and metallicity ranges used from each work. To integrate these datasets, we create a look-up ``net stellar yield" table that extends over the mass range $1-8$ M$_{\odot}$ and metallicity range $Z=10^{-4}$ to $Z=0.1$. We expand the net yields from \citet{Karakas10} and \citet{Fishlock14} to the mass range $6-8$ M$_{\odot}$ by linearly extrapolating the yields in mass. In the case of the $Z=0.004$ net yields, we extend the \citet{Karakas10} data by incorporating the net yields from \citet{Doherty14}. In the lookup table, both the input parameters\textemdash initial mass and metallicity\textemdash and the output\textemdash net yields\textemdash are on a linear scale. 

To calculate the mass of element $k$ ejected by a single star, $m_{{\rm{ej}},k}$, we use net stellar yield tables, $Y_{k,\rm{tbl}}(M_{\rm{tbl}},Z_{\rm{tbl}})$ (in solar mass units), which give the total mass of element $k$ that a star of zero-age main sequence (ZAMS) mass $M_{\rm{tbl}}$ and metallicity $Z_{\rm{tbl}}$ produces and ejects into the ISM during its entire lifetime. As a result, $Y_{k,\rm{tbl}}$ does not include the initial mass of element $k$ that was present in the star. It is possible that $Y_{k,\rm{tbl}}$ takes negative values. In that case, it reflects the quantity of species $k$ that is consumed inside the star over its lifetime. Conversely, positive yields quantify the production of species $k$ over the star's lifetime.

For clarity, throughout this section we use the subscript ``tbl'' to denote tabulated values taken from stellar yields tables, while the subscript ``sim'' indicates values used and predicted by the simulations.

We calculate $m_{{\rm{ej}},k}$ as follows

\begin{eqnarray}\label{mej_method_AGB}
m_{{\rm{ej}},k}(M_{\rm{sim}}, Z_{i,\rm{sim}}) &=& Y_{k,\rm{tbl}}(M_{\rm{sim}}, Z_{i,\rm{sim}}) +\\\nonumber
&& X_{i,k,\rm{sim}}m_{\rm{ej,tbl}}(M_{\rm{sim}}, Z_{i,\rm{sim}}),
\end{eqnarray}

\noindent where, given the specific ZAMS star mass $M_{\rm{sim}}$ and metallicity $Z_{i,\rm{sim}}$ of the star particle $i$, $Y_{k,\rm{tbl}}$ is extracted by linearly interpolating (in log-log space) the yields tables. Note that $M_{\rm{sim}}$ is not the mass of the star particle $i$, but rather a ZAMS star mass from the SSP the star particle represents (see eq.~\ref{deltami}).

In eq.~(\ref{mej_method_AGB}), we also include the mass that is present in the star by adding the product of the abundance of element $k$, $X_{i,k,\rm{sim}}$, with the tabulated total mass ejected by the star during its lifetime $m_{\rm{ej,tbl}}$, giving as a result the total mass ejected (both initially present in the star and produced/destroyed by stellar nucleosynthesis). 

To determine $m_{\rm{ej,metal}}$, we also apply eq.~(\ref{mej_method_AGB}), use $Y_{\rm{metal},\rm{tbl}}(M_{\rm{sim}}, Z_{i,\rm{sim}})$ from the yield tables, and replace $X_{i,k,\rm{sim}}$ by $Z_{i,\rm{sim}}$. Finally, to calculate $m_{\rm{ej,total}}$, we directly use $m_{\rm{ej,tbl}}(M_{\rm{sim}}, Z_{i,\rm{sim}})$ from the yield tables.

In summary, the stellar yield tables include star masses, $M_{\rm{tbl}}$, and metallicities, $Z_{\rm{tbl}}$, net yields, $Y_{k,\rm{tbl}}(M_{\rm{tbl}},Z_{\rm{tbl}})$ for each element $k$ that we track as well as for total metals, and the total mass ejected, $m_{\rm{ej,tbl}}(M_{\rm{tbl}},Z_{\rm{tbl}})$.

\subsubsection{$s$-process nucleosynthesis}

AGB stars serve as significant sources of heavy elements beyond iron, predominantly through the slow neutron capture process, or $s$-process (e.g. \citealt{Sneden08}). \citet{Karakas16} present $s$-process yields specifically for barium and strontium. \citet{Kobayashi20} indicate that the over-production of the second (Ba) and third (Pb) $s$-process peak elements at [Fe/H] = 0 means that the contribution of $s$-process elements is overestimated by \citet{Karakas16}. They propose reducing this contribution by decreasing the parameter $M_{\rm{mix}}$, which regulates the extent in mass of the region where $s$-process elements are produced. Following this comment, we opt to decrease the \citet{Karakas16} net yields of barium by a factor of 2. 

\citet{Fishlock14} also provide $s$-process yields, but their net yields are lower than those from \citet{Karakas16}. If incorporated in our models, this would result in a declining efficiency of $s$-process elements production in our model with decreasing metallicity. However, this would contradict observations of low-metallicity stars rich in Sr (see e.g. \citealt{Roederer14,Zhao16, Spite18}). Consequently, we do not use the $s$-process yields from Fishlock et al. Instead, we linearly extrapolate (in log-log space) the $s$-process net yields from \citet{Karakas16} to the metallicity bins $Z=10^{-4}$, 0.001 and 0.004, favoring yields rich in $s$-process elements in low-metallicity stars.

\citet{Cinquegrana22} and \citet{Doherty14} provide datasets that lack individual yields for Ba and Sr, instead they provide an $s$-process tracer ($g$) that estimates the amount of material heavier than Fe produced through the AGB phase. We use the $g$ tracer from \citet{Cinquegrana22} and assume that a fraction of $10^{-3}$ corresponds to strontium and a fraction of $2\times 10^{-4}$ corresponds to barium. These factors give net yields of the order of ${\sim}10^{-9}~\rm{M}_{\odot}$ for Sr and ${\sim}10^{-10}~\rm{M}_{\odot}$ per star for Ba, aligning with yields from other studies (e.g. \citealt{Cristallo16, Prantzos18, Limongi18}) and slightly below the lowest $s$-process net yields from the highest metallicity bin from \citet{Karakas16}. It is important to note the large uncertainty associated with $s$-process yields due to various factors, including neutron flux, neutron-to-seed mass ratio, the source (e.g. the nuclear reaction $^{13}{\rm{C}}(\alpha,n)^{16}{\rm{O}}$ for low-mass stars and $^{22}{\rm{Ne}}(\alpha,n)^{25}{\rm{Mg}}$ for massive stars) and the inter-shell mixing from stellar rotation (see e.g. \citealt{Pignatari08,Chiappini11,Frischknecht16}).

The calculation of $m_{\rm{ej,Ba}}$ and $m_{\rm{ej,Sr}}$ from the net yields follows the methodology for general AGB yields described in the previous subsection.

\subsubsection{$r$-process nucleosynthesis}\label{r_process_sec}

The Advanced LIGO/Virgo detection of a neutron star merger, GW170817, coupled with the subsequent observations across ultraviolet, optical and infrared wavelengths, has provided compelling evidence for $r$-process nucleosynthesis (see e.g. \citealt{Abbott17c,Abbott17b}). Many low-metallicity stars exhibit high $r$-process abundances compared to iron, surpassing solar values by up to two orders of magnitude (e.g. \citealt{Roederer18,Hansen18,Cain20,Aguado21,Reggiani21}). Such enhancements pose challenges to scenarios where NSMs are the sole source of $r$-process elements (e.g. \citealt{Wehmeyer15,Cote19,vandeVoort22,Kobayashi23b}), as the ISM hosting the NSM event should already be enriched with iron from the CCSNe that formed the neutron stars. 

We explore alternative astrophysical sites for $r$-process nucleosynthesis, including CEJSN and collapsars. \citet{Papish15}, along with \citet{Grichener19}, propose that jets launched by a neutron star during a CEJSN explosion can serve as a site for heavy $r$-process nucleosynthesis. In this scenario, the CEJSN event begins with a detached binary system of two massive main sequence stars. The more massive star evolves to a red super-giant, eventually exploding to leave behind a neutron star remnant. Once the (initially less massive) star expands to become a red super-giant, it swallows the neutron star, setting the system on course to evolve towards a CEJSN. During this phase, the neutron star penetrates the envelope of the giant star after the exhaustion of helium in its core. The accretion rate onto the neutron star is sufficiently high for heavy $r$-process nucleosynthesis to occur (e.g. \citealt{Grichener22}).

Collapsars, resulting from the core-collapse of a massive star forming a rotating BH and an accretion disk, present another potential site for $r$-process nucleosynthesis (e.g. \citealt{Brauer21,Barnes22}). \citet{Siegel19} propose that collapsar accretion disks generate neutron-rich outflows synthesizing heavy $r$-process nuclei. Collapsars occur primarily in low-metallicity environments, and thus provide a natural explanation for the observed carbon-enhanced metal-poor stars with high $r$-process enrichment. However, debate persists regarding the efficiency of collapsars as $r$-process producers (e.g. \citealt{Miller20, Just22, Fujibayashi23, Dean24}). 

Our study investigates europium abundances in scenarios where NSM, CEJSN, and collapsars are the exclusive sources of $r$-process elements.

We also considered magneto-rotational SNe (e.g. \citealt{Winteler12,Reichert21}) and neutrino winds from CCSNe (e.g. \citealt{Roberts17}) as potential $r$-process production sites. However, we discarded the magneto-rotational SNe scenario based on the unrealistically large magnetic fields needed in the precollapse source to produce heavy $r$-processes (\citealt{Mosta18}). For the case of neutrino winds from CCSNe, while it has been suggested that winds from CCSNe coming from regions close to the neutron star should be neutron-rich, studies of neutrino radiation hydrodynamics simulations (e.g. \citealt{Fischer18}) have shown that the proto-neutron star losses leptons, it emits neutrinos of all flavors, and produces ejecta that are purely proton-rich (e.g. \citealt{Pinedo14,Pinedo17}). We therefore did not consider this scenario either. 

\begin{table}
\caption{List of some of the properties assumed for the three $r$-process astrophysical production sites considered in this work: neutron star mergers (NSM), common envelope jets SNe (CEJSN) and collapsars. We elaborate on these properties in Section~\ref{r_process_sec}. References: $^{a}$Adv. LIGO/Virgo (\citealt{Abbott17a}); $^b$Type II-P SN (\citealt{Arcavi17}); $^c$Fast-rising blue optical transient (\citealt{Prentice18}); $^d$long Swift gamma-ray bursts (\citealt{Wanderman10}).}
\begin{center}
\begin{tabular}{ l|c|c|c }
\hline
& NSM & CEJSN & Collapsar \\
\hline\hline
Observational & GW170817$^{a}$ & iPTF14hls$^{b}$ & Long-duration \\
signature &   & AT2018cow$^{c}$ & GRBs$^{d}$ \\
\hline
Cosmic rate & 1540$^{+3200}_{-1220}$  & $f{\times}7.05^{+1.43}_{-1.25}{\times}10^{4}$& 1.3$^{+0.6}_{-0.7}/f_{b}$ \\
$[{\rm{Gpc}}^{-3}~\rm{yr}^{-1}]$ & & $f{\sim}1\%$ & $f_{b}\approx 5{\times}10^{-3}$ \\
\hline
Normalization & $1.3\times 10^{-5}$ & $1.22\times 10^{-5}$ &  $2.6\times 10^{-5}$\\
$[\rm{M}_{\odot}^{-1}]$ & & & \\
\hline
Delay time & 30 & 0 & 0 \\
 $t_{\rm{min}}$ [Myr] & & & \\
\hline
Eu Yields & $6\times 10^{-6}$ & $2\times 10^{-5}$ & $10^{-6}$\\
($Y_{\rm{Eu,tbl}}$) [M$_{\odot}$] & & & \\
\hline
\end{tabular}
\end{center}
\label{r_process_table}
\end{table}

The calculation of the mass of europium ($\Delta m_{i,\rm{Eu}}$) ejected during a single time step ($\Delta t$) by a star particle $i$ differs from the method described in eqs.~(\ref{deltami}) and (\ref{mej_method_AGB}). Instead, we first determine the number of events from NSM, CEJSN, and collapsars, $N_{i,\rm{NSM/CEJSN/coll}}$, that occur during each $\Delta t$ in the SSP represented by the star particle. This is done by calculating $N_{i,\rm{NSM/CEJSN/coll}}$ based on the rate of events, $R_{i,\rm{NSM/CEJSN/coll}}$, per unit time, and integrating over $\Delta t = t_{j+1} - t_{j}$, as follows

$$N_{i,\rm{NSM/CEJSN/coll}}=\int_{t_{j}}^{t_{j+1}}R_{i,\rm{NSM/CEJSN/coll}}(t){\rm{d}}t.$$ 

\noindent From this $\Delta m_{i,\rm{Eu}}$ is calculated as follows, 

\begin{eqnarray}
\Delta m_{i,\rm{Eu}} &=& N_{i,\rm{NSM}}Y_{\rm{Eu,NSM,tbl}} + \\\nonumber
& & N_{i,\rm{CEJSN}}Y_{\rm{Eu,CEJSN,tbl}} + N_{i,\rm{coll}}Y_{\rm{Eu,coll,tbl}},
\end{eqnarray}

\noindent where $Y_{\rm{Eu,NSM/CEJSN/coll,tbl}}$ corresponds to the tabulated yields of europium (in units of solar mass) from NSM, CEJSN and collapsars that we adopt. Once $\Delta m_{i,\rm{Eu}}$ is known, the distribution of the europium mass to the neighboring gas particles follows eq.~(\ref{deltami2}).

We parametrize the rate of NSM events produced by an SSP, $R_{i,\rm{NSM}}$, per unit time, as 

\begin{equation}\label{R_NSM}
R_{i,\rm{NSM}}=c_{\rm{NSM}}m_{i}\Theta(t-t_{\rm{delay}})t^{-1},
\end{equation}

\noindent where $c_{\rm{NSM}}$ is the normalization per unit mass of NSM events, $m_{i}$ is the initial mass of the star particle $i$, $\Theta(t-t_{\rm{delay}})$ is the Heaviside step function representing the delay time distribution function with $t_{\rm{delay}}$ the minimum delay time of 30 Myr. 

The rates of CEJSN and collapsar events, $R_{\rm{CEJSN}}$ and $R_{\rm{coll}}$ respectively, are assumed to be

\begin{eqnarray}\label{R_CEJSN}
R_{i,\rm{CEJSN}} &=& c_{\rm{CEJSN}}m_{i} [\Delta\tau(\Delta M,Z_{i})]^{-1},\label{R_coll}\\
R_{i,\rm{coll}} &=& c_{\rm{coll}}m_{i} [\Delta\tau(\Delta M,Z_{i})]^{-1},
\end{eqnarray}

\noindent where $\Delta\tau(\Delta M,Z_{i})=\tau(M_{\rm{min}},Z_{i})-\tau(M_{\rm{max}},Z_{i})$ is the time range over the stars' lifetimes with $M_{\rm{min}}=$ 8 M$_{\odot}$ and $M_{\rm{max}}=$ 40 M$_{\odot}$ for CEJSN and $M_{\rm{min}}=$ 10 M$_{\odot}$ and $M_{\rm{max}}=$ 40 M$_{\odot}$ for collapsars. The normalization constants, $c_{\rm{CEJSN}}$ and $c_{\rm{coll}}$, correspond to the local observed rates per unit mass.

We adjust the normalizations, so that when we integrate $R_{i,\rm{NSM}}$, $R_{i,\rm{CEJSN}}$ and $R_{i,\rm{coll}}$ over all star particles within a given cosmic volume at $z=0$, we reproduce the observed present-day cosmic rates of NSM (\citealt{Abbott17a}), CEJSN (\citealt{Li11,Chevalier12}) and collapsars (\citealt{Wanderman10}). The cosmic rates, normalizations and additional properties are listed in Table~\ref{r_process_table}. For CEJSN, we follow \citet{Chevalier12}, who estimated that the rate of events in which a NS enters the envelope of a giant is about $1\%$ of the rate of CCSNe. For collapsars, we adopt the local (redshift $z = 0$) volumetric rate of classical long Gamma-Ray Bursts (GRBs) of $\approx 0.6-2$ Gpc$^{-3}$ yr$^{-1}$ (\citealt{Wanderman10}), and a gamma-ray beaming fraction of $f_{\rm{b}} = 0.005$ (\citealt{Goldstein16}). Note that for NSM events the maximum rate of events per star particle after 30 Myr is $R_{i,\rm{NSM}}\approx 5\times 10^{-13}$ yr$^{-1}$ M$_{\odot}^{-1}$. For a particle mass resolution of $10^{6}~\rm{M}_{\odot}$ ($10^{4}~\rm{M}_{\odot}$), the maximum number of NSM events a star particle can produce in a timestep $\Delta t$ is $5\times 10^{-7}(\Delta t/$yr$^{-1}$) ($5\times 10^{-9}(\Delta t/$yr$^{-1}$)). Given these low numbers of events per time step (the same holds for CEJSN and collapsars), we implement the $r$-process enrichment in a stochastic manner. 

The europium yields for NSM, CEJSN, and collapsars are listed in Table~\ref{r_process_table}. For NSM, we use the yields reported by \citet{Cote18}, who found that to match the observed europium levels in the Milky Way, each NSM event must eject an average of $3\times 10^{-6}~\rm{M}_{\odot}$ to $10^{-5}~\rm{M}_{\odot}$ of europium. These values align with the estimated ejecta mass from GW170817 and are consistent with results from both chemical evolution studies and population synthesis. For CEJSN, we adopt the yields reported by \citet{Grichener19}, who used the MESA stellar evolution code to follow the evolution of low-metallicity giant stars that swallow neutron stars during their late expansion phases. For collapsars, we use the yields from \citet{Siegel19}, who conducted general-relativistic magnetohydrodynamics simulations of collapsing BHs and performed nuclear reaction network calculations on tracer particles using the code SkyNet.

\subsubsection{CCSN yields}

The calculation of the total mass, metal mass, and the mass of element $k$ ($\Delta m_{i,{\rm{total/metal}}/k}$) ejected in each time step by a star particle $i$ due to CCSNe enrichment follows eqs.~(\ref{deltami}) and (\ref{deltami2}). As with AGB enrichment, the values for $m_{{\rm{ej}},{\rm{total/metal}}/k}$ are taken from nucleosynthesis tables.

We calculate $m_{{\rm{ej}},k}$ as follows

\begin{eqnarray}\label{mej_method_CCSN}
m_{{\rm{ej}},k}(M_{\rm{sim}}, Z_{i,\rm{sim}}) &=& m_{{\rm{ej,tbl/CCSN}},k}(M_{\rm{sim}}, Z_{i,\rm{sim}}) +\\\nonumber
&& X_{i,k,\rm{sim}}m_{\rm{ej,tbl/winds}}(M_{\rm{sim}}, Z_{i,\rm{sim}}),
\end{eqnarray}

\noindent where $m_{{\rm{ej,tbl/CCSN}},k}$ is the mass of element $k$ produced and ejected into the ISM by the supernova explosion, and $m_{\rm{ej,tbl/winds}}$ corresponds to the total mass ejected into the ISM via stellar winds during the pre-supernova phase of the star. We calculate $m_{\rm{ej,tbl/winds}}=M_{\rm{sim}}-M_{\rm{final}}(M_{\rm{sim}}, Z_{i,\rm{sim}})$, with $M_{\rm{sim}}$ the zero-age main sequence mass of the star and $M_{\rm{final}}$ the pre-supernova phase mass. In the above equation, we have assumed that the pre-SN ejecta have the same composition as the star had at birth. We remind the reader that, similar to eq.~(\ref{mej_method_AGB}), $M_{\rm{sim}}$ is not the mass of the star particle $i$, but rather a zero-age star mass from the SSP the star particle represents. Also, that the subscript “tbl” denotes tabulated values taken from stellar yields tables, whereas “sim” indicates values from the simulations.

An approximation in our treatment, given in eq.~(\ref{mej_method_CCSN}), of the pre-supernova wind ejecta combined with the supernova ejecta is that both components are released simultaneously. While this assumption is not strictly accurate\textemdash since massive stars, particularly in the red supergiant or Wolf-Rayet phase, can undergo significant mass loss via stellar winds over timescales ranging from 10,000 years to a few million years\textemdash we adopt it here for the sake of simplicity.

Following a similar approach to that used for AGB yields, we generate look-up tables containing $m_{{\rm{ej,tbl/CCSN}},k}(M_{\rm{tbl}}, Z_{\rm{tbl}})$, $m_{\rm{ej,tbl/winds}}(M_{\rm{tbl}}, Z_{\rm{tbl}})$, and the total mass ejected by the star, $m_{\rm{ej,tbl}}(M_{\rm{tbl}}, Z_{\rm{tbl}})$, during both the pre-supernova phase and the CCSN explosion, for stars within the ZAMS mass range $6-40~\rm{M}{\odot}$. To calculate the total ejected mass, $m_{{\rm{ej}},{\rm{total}}}$, we directly use $m_{\rm{ej,tbl}}$ from these tables. The total ejected metal mass, $m_{\rm{ej},{\rm{metal}}}$, is then determined by $m_{\rm{ej},{\rm{metal}}} = m_{\rm{ej},{\rm{total}}} - m_{\rm{ej,H}} - m_{\rm{ej,He}}$, where $m_{\rm{ej,H}}$ and $m_{\rm{ej,He}}$ are computed using eq.~(\ref{mej_method_CCSN}). This ensures that we track all metals produced by the CCSN event, regardless of whether they are individually traced or not.

We adopt the CCSN nucleosynthesis yields from the tables by \citet{Nomoto13}. These tables comprise datasets generated by \citet{Kobayashi06} and \citet{Nomoto06}, presented as a function of the progenitor (ZAMS) mass ($M_{\rm{tbl}}=$ 13, 15, 18, 20, 25, 30, and 40 $\rm{M}_{\odot}$) and metallicity ($Z_{\rm{tbl}}=$ 0, 0.001, 0.004, 0.02, and 0.05). It is important to note that these yields specifically account for the metals ejected during the nucleosynthesis process of the CCSN ejecta. They do not include metals expelled into the ISM through stellar winds in the pre-supernova phase of the stars. To address this limitation, we calculate the pre-supernova stellar mass loss by following the work of \citet{Kobayashi06}, which gives $M_{\rm{final}}(M_{\rm{tbl}},Z_{\rm{tbl}})$, the total mass lost through stellar winds in the pre-supernovae phase. Note that, since the tables of \citet{Nomoto13} cover a mass range starting from $13~\rm{M}_{\odot}$, we extend our data by linearly extrapolating below this value. In our fiducial model, stars within the mass range of $8-40~\rm{M}_{\odot}$ evolve into CCSNe, contributing to the enrichment of the ISM. Stars exceeding $40~\rm{M}_{\odot}$ are assumed to become faint supernovae, re-accreting the bulk of their material onto their remnant BHs.

\citet{Nomoto13} provide nucleosynthetic yields not only for CCSNe but also for hypernovae (HNe), a subset of CCSNe characterized by significantly higher explosion energies ($E_{51} \equiv E/10^{51}{\rm{erg}} > 10$). The primary distinction between CCSNe and HNe in their models lies in the assumed explosion energy used for yield calculations: \citet{Nomoto13} assume an explosion energy of $10^{51}{\rm{erg}}$ for CCSNe, while for HNe, they employ a mass-energy relation. In this work, we adopt only the yields for CCSNe.

\citet{Wiersma09} noted, based on a survey of the literature, that relative abundance ratios derived from nucleosynthesis yields may only be reliable at the two-factor level, even for a fixed IMF. In light of this, we use the established approach of introducing ``boost factors'' as free parameters, calibrated to help reproduce various stellar abundance ratios when comparing our findings with the APOGEE dataset (\citealt{Holtzman18}), among others (see Sec.~\ref{CCSN_boost_factors_section}). In our fiducial model, the boost factors, denoted as $f_{k}$ (with $k$ representing the element), are $f_{k}=1.0$ for all elements except carbon and magnesium, for which $f_{\rm{C}}=f_{\rm{Mg}}=1.5$. The total mass of a specific element $k$, $m_{{\rm{ej}},k}$ (eq.~\ref{mej_method_CCSN}), produced in a timestep is then multiplied by the corresponding $f_{k}$. This means that the total ejected mass and the total ejected metal mass increase in accordance with the boosted yields. 

\subsubsection{SNIa}

The progenitor systems of SNe Ia are still debated. Possible scenarios include delayed detonations of Chandrasekhar- (Ch-) mass white dwarfs (WDs) from single or double degenerate systems, or double detonations of sub-Ch-mass WDs (e.g., \citealt{Liu23} for a review). In this work, we use the yields derived by \citet{Leung18}, who created several SNIa models for near-Ch mass WDs using the turbulent deflagration model with deflagration-detonation transition. We adopt their W7 model, a recalculated version of the model from \citet{Iwamoto99}, which incorporates updated electron capture and nuclear reaction rates. This model has been shown to better reproduce the Cr/Fe ratios of the Perseus cluster than a variety of other SN Ia nucleosynthesis models (e.g. \citealt{Pakmor12, Seitenzahl13, Shen18}) that  match the properties of typical SN Ia explosions (e.g., the mass of $^{56}$Ni and hence the maximum brightness) most closely (\citealt{Simionescu19}).

To determine the number of SNIa explosions per timestep, we sample the SNIa rate from a delay time distribution function (DTD). The DTD is defined as DTD$(t)=\nu\xi(t)$, where $\nu$ is the total number of SNIa per unit initial stellar mass and $\xi(t)$ is a function normalised to 1 over the age of the Universe. In our fiducial model we parametrise the DTD using an exponential function, $\xi(t)={\rm{exp}}(-t/\tau)$, where $\tau$ is a free parameter. We also assume that SNIa have a fixed minimum delay time of $t_{\rm{delay}}=40$ Myr corresponding to the time elapsed between the formation of the progenitor star and the explosion of the SNIa. The DTD then becomes

\begin{equation}
{\rm{DTD}}(t) = \frac{\nu}{\tau}\frac{{\rm{exp}}(-t/\tau)\times\Theta(t-t_{\rm{delay}})}{[{\rm{exp}}(-t_{\rm{delay}}/\tau)-{\rm{exp}}(-t_{\rm{H}}/\tau)]},
\end{equation}

\noindent where $\nu=1.54 \times 10^{-3} \rm{M}_{\odot}^{-1}$, $\tau=2$ Gyr and $t_{\rm{H}}$ is the Hubble time.

Other DTD functions have also been assumed in the literature, such as a power law, i.e. $\xi(t)\propto t^{-\beta}$, with $\beta=1$ inferred from the cosmic SNIa rate in both the field and clusters (\citealt{Maoz12a, Maoz14}), and a gaussian function, i.e. $\xi(t)\propto {\rm{exp}}(-[(t-\tau)/2\sigma]^{2})$ (e.g. \citealt{Dahlen04}). In the work of Nobels et al. (in prep.), we investigate three DTD models: exponential, power-law, and Gaussian. We find that the observed cosmic SNIa rate can be reproduced using all DTDs, provided the SNIa efficiency and average delay time are adjusted. Galaxy metal abundances, in particular the alpha-to-iron ratios as a function of the iron abundance, favour somewhat longer delay times, irrespective of the DTD assumed.

\begin{figure*} 
\begin{center}
\includegraphics[angle=0,width=\textwidth]{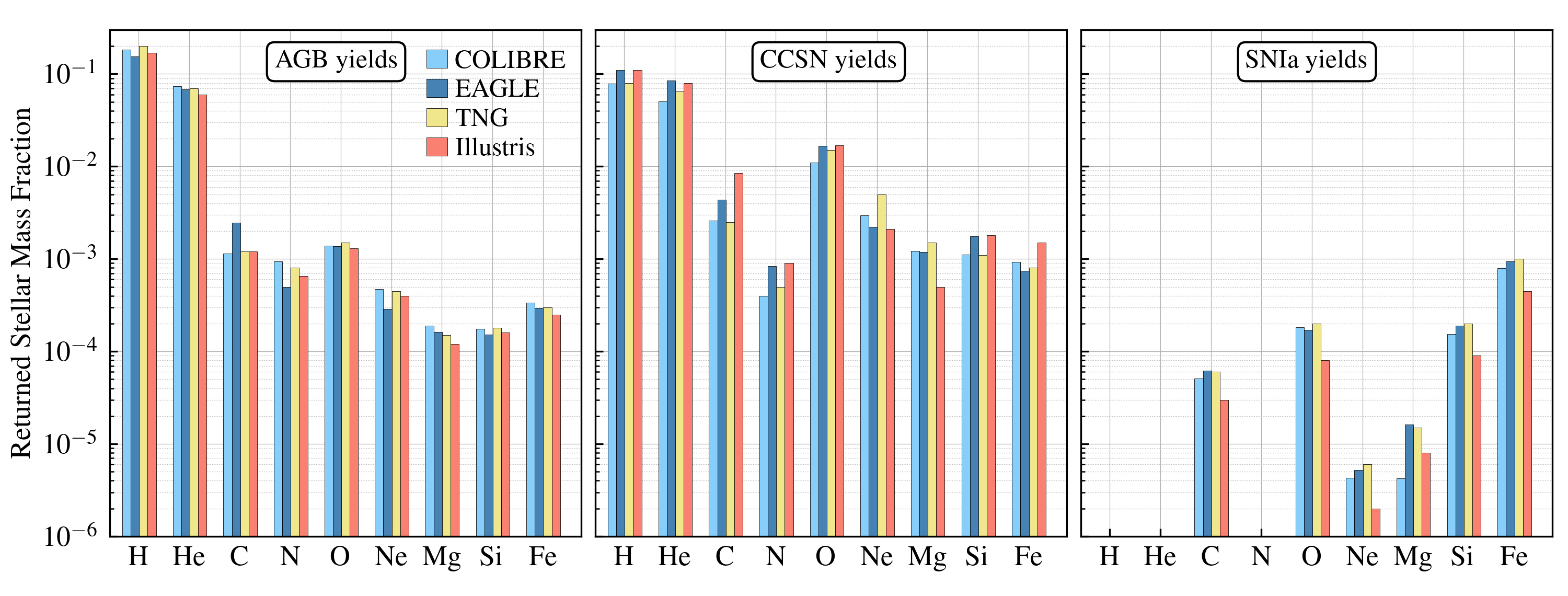}\\
\caption{Fraction of mass returned to the ISM in a Hubble time per unit stellar mass formed at Solar metallicity (Z$_{\odot}=0.0134$, \citealt{Asplund09}). The figure compares the mass returned by AGB (left panel), CCSN and winds from their progenitors (middle panel), and SNIa (right panel) in the COLIBRE, EAGLE, TNG and Illustris galaxy formation models.}
\label{ComparisonYields}
\end{center}
\end{figure*}

\subsection{Yields comparison}\label{Sec22}

We compare the set of yield tables adopted in this work with others from the literature. To do so we calculate the stellar mass fraction by chemical species that is returned to the ISM in a Hubble time, $t_{\rm{H}}$, per unit stellar mass formed, at solar metallicity. We extract from the yield tables, $m_{{\rm{ej}},k}(M,{\rm{Z}}_{\odot})$ (of a given element $k$), that is ejected into the ISM by a star of ZAMS mass $M$ and metallicity ${\rm{Z}}_{\odot}$ over its lifetime. We integrate $m_{{\rm{ej}},k}$ over the IMF, $\phi(M)$, and normalize by the integral of the star masses over the IMF. Specifically, the returned stellar mass fraction, RSMF$_{k}$, of element $k$, is calculated as

\begin{equation}
{\rm{RSMF}}_{k,{\rm{AGB/CCSN}}}({\rm{Z}}_{\odot}) = \frac{\int_{{\rm{min}}(M_{Z}(t<t_{\rm{H}}))}^{100~{\rm{M}}_{\odot}}  \phi(M)\times m_{{\rm{ej}},k}(M,{\rm{Z}}_{\odot}) {\rm{d}}M}{\int_{0.1~{\rm{M}}_{\odot}}^{100~{\rm{M}}_{\odot}} \phi(M)\times M{\rm{d}}M},
\end{equation}

\noindent where the numerator corresponds to the total mass of element $k$ ejected by stars (via the AGB or CCSN phase) with lifetimes shorter than a Hubble time ($M_{Z}(t<t_{\rm{H}})$). 

The stellar mass that is returned to the ISM by Type Ia supernovae depends on the number of SNIa events per unit stellar mass over a Hubble time, $N_{\rm{SNIa}} =\int_{t_{\rm{delay}}}^{t_{\rm{H}}}{\rm{DTD}}(t'){\rm{d}}t' = \nu$. The stellar mass fraction ejected into the ISM by SNIa is then

\begin{equation}
{\rm{RSMF}}_{k,\rm{SNIa}} = \frac{\int_{{\rm{min}}(M_{Z}(t<t_{\rm{H}}))}^{100~{\rm{M}}_{\odot}}  \phi(M)\times N_{\rm{SNIa}}\times M \times m_{{\rm{ej}},k}{\rm{d}}M}{\int_{0.1~{\rm{M}}_{\odot}}^{100~{\rm{M}}_{\odot}} \phi(M)\times M{\rm{d}}M}.
\end{equation}

Fig.~\ref{ComparisonYields} shows the fraction of mass returned to the ISM by AGB stars (left panel), CCSN (middle panel) and SN Type Ia (right panel), and compares the fractions computed using the yield tables adopted in this work with those from the EAGLE (\citealt{Schaye15}), TNG (\citealt{Pillepich18}) and Illustris (\citealt{Vogelsberger14}) galaxy formation models. These last two are extracted from fig. 1 of \citet{Pillepich18}. The EAGLE model follows the enrichment implementation from the OWLS simulations (\citealt{Wiersma09,Schaye10}). For AGB and CCSN it uses the nucleosynthesis yields from \citet{Marigo01} and \citet{Portinari98}, respectively, as well as the SNIa yields of the W7 model of \citet{Thielemann03}. TNG adopts yields that are somewhat similar to the ones used in this work. For AGB, it joins the yields tables from \citet{Karakas10}, \citet{Doherty14} and \citet{Fishlock14}, while for CCSN it uses the yield tables of \citet{Nomoto13} over the mass range $13-40$ M$_{\odot}$, but it extends them to the mass range $8-120$ M$_{\odot}$ by interpolating the yields of \citet{Portinari98}. For SNIa, TNG adopts the yields from \citet[model W7]{Nomoto97}. Finally, the Illustris model adopts a simpler version of the TNG yield tables, it only uses the yield tables from \citet{Karakas10} for AGB enrichment, from \citet{Portinari98} for CCSN and joins the tables from \citet{Travaglio04} and \citet{Thielemann03} for SNIa. 

Note that both EAGLE and COLIBRE adopt boost factors ($f_{k}$ for element $k$) for CCSN enrichment. This means that the total mass of a specific element $k$, $m_{{\rm{ej}},k}$, produced in a timestep is multiplied by $f_{k}$. In COLIBRE the boost factors are $f_{\rm{C}}=1.5$ and $f_{\rm{Mg}}=1.5$ for carbon and magnesium, and $f_{k}=1$ for the remainder of the elements. In contrast, EAGLE uses the following boost factors: $f_{\rm{H}}=1.0$, $f_{\rm{He}}=1.0$, $f_{\rm{C}}=0.5$, $f_{\rm{N}}=1.0$, $f_{\rm{O}}=1.0$, $f_{\rm{Ne}}=1.0$, $f_{\rm{Mg}}=2$, $f_{\rm{Si}}=1.0$, and $f_{\rm{Fe}}=0.5$. These boost factors are taken into account when we calculate the ${\rm{RSMF}}_{k,{\rm{CCSN}}}$. TNG does not apply boost factors. Instead, when TNG combines the CCSN yields from \citet{Nomoto13} and \citet{Portinari98}, the models are renormalized so that the IMF-weighted yield ratios at each metallicity match those derived from the Nomoto mass range models alone.

For AGB stars, the COLIBRE yields are comparable to those of TNG, differing by at most a factor of 1.2, as expected, since both are based on the Karakas et al. nucleosynthesis models. Similarly, COLIBRE and Illustris yields are generally consistent, except for nitrogen, magnesium, and iron, where COLIBRE yields are higher by factors of 1.35 to 1.6. In contrast, EAGLE consistently produces lower AGB yields than COLIBRE, particularly for nitrogen (by a factor of ${\approx}1.9$) and neon (by ${\approx}1.65$), while for carbon, EAGLE yields are higher by a factor of ${\approx}2.1$. Differences are also evident in the yields from CCSNe. COLIBRE yields are generally similar to TNG, except for nitrogen, oxygen, neon, and magnesium, where COLIBRE yields are lower by factors of 1.25 to 1.7. This discrepancy is somewhat expected, as COLIBRE does not extend CCSNe yields beyond the $8-40$ M$_{\odot}$ mass range, unlike TNG. EAGLE and Illustris adopt the same CCSNe yields; however, differences for elements such as carbon, magnesium, and iron may arise from the boost factors (0.5, 2.0, and 0.5, respectively) employed by EAGLE. For SNIa, the yields of carbon, oxygen, silicon, and magnesium are consistent within a factor of 1.2 among COLIBRE, EAGLE, and TNG. However, for neon, TNG yields are a factor of 1.4 higher than those of COLIBRE, and for magnesium, COLIBRE yields are a factor of 3.5 lower than those of EAGLE and TNG. Illustris, in general, shows lower SNIa yields for all elements.

\subsection{Metal diffusion}\label{Metal_diffusion_Sec}

The mixing of enriched material ejected by stars with surrounding gas is important for the overall matter cycle in galaxies. The rate at which metals mix in the ISM can depend on the energy of the enrichment source (e.g. \citealt{Emerick18, Emerick20}), and it affects the cooling rate of the gas as well as the metal composition of new stars. In galaxy formation simulations, accurately modeling this mixing is challenging because the turbulent motions that are responsible for the mixing occur at scales too small to resolve. To address this, small-scale chemical mixing is often modeled as a diffusion process (e.g. \citealt{Martinez08, Greif09,Shen10}), $\frac{{\rm{d}}X_{i}}{{\rm{d}}t}=\frac{1}{\rho_{\rm{g}}}\nabla\cdot(D\nabla X_{i})$, where $X_{i}$ is the species abundance of gas particle $i$, $\rho_{\rm{g}}$ the gas density and $D$ the diffusion coefficient. The time derivative is Lagrangian. In the SPH formalism the diffusion equation can be reduced to a discrete summation over all particles $j$ within the smoothing kernel of gas particle $i$,

\begin{equation}\label{2}
\frac{{\rm{d}}X_{i}}{{\rm{d}}t}=\sum_{j}^{N_{\rm{ngb}}}K_{ij}(X_{i}-X_{j}),
\end{equation}

\noindent with

\begin{equation}\label{3}
K_{ij}=\frac{m_{j}}{\rho_{i}\rho_{j}}\frac{4D_{i}D_{j}}{D_{i}+D_{j}}\frac{{\bf{r}}_{ij}\cdot \nabla_{i}\bar{W}_{ij}(r_{ij},h_{i})}{r_{ij}^{2}},
\end{equation}

\noindent where $i$ and $j$ denote the particle indices, $m$ the mass, $\rho$ the density, $\bar{W}_{ij}$ the average kernel (dimension length$^{-3}$) and $r_{ij}$ the vector and absolute separations between particles $i$ and $j$, respectively (\citealt{Monaghan05}). As a result, $K_{ij}$ has units of time$^{-1}$. Note that in eq.~(\ref{3}) we are using the average of the kernel to maintain the symmetry in the diffusion equation (e.g. \citealt{Monaghan92}), that is

\begin{equation}
\bar{W}_{ij}=\frac{1}{2}[W_{ij}(r_{ij},h_{i})+W_{ji}(r_{ij},h_{j})],
\end{equation}

\noindent and correspondingly,

\begin{equation}
\nabla_{i}\bar{W}(r_{ij},h_{i}) = \frac{1}{2}[\nabla_{i}W_{ij}(r_{ij},h_{i})+\nabla_{i}W_{ji}(r_{ij},h_{j})].
\end{equation}

We can integrate eq.~(\ref{2}) using an explicit Euler time stepping scheme, advancing from timestep $n$ to $n+1$. This yields the abundance of element $X$ for particle $i$ at timestep $n+1$ as

\begin{equation}\label{4}
X_{i}^{n+1} = X_{i}^{n} + \sum_{j}^{N_{\rm{ngb}}}K_{ij}(X^{n}_{i}-X^{n}_{j})\Delta t_{ij}.
\end{equation}

\noindent To conserve metals, so that the exchange of metal mass between particles $i$ and $j$ is $\Delta X_{i}m_{i} = -\Delta X_{j}m_{j}$, where $\Delta X_{i/j}=X^{n+1}_{i/j}-X^{n}_{i/j}$ (i.e. the change in abundance due to diffusion), we make sure that $m_{i}K_{ij}=m_{j}K_{ji}$ (through the symmetry of the kernel), and that $\Delta t_{ij}=\Delta t_{ji}$. However, since different active particles may have different time steps, we impose $\Delta t_{ij}=\Delta t_{ji}={\rm{min}}(\Delta t_{i},\Delta t_{j})$ and use the shorter time step in the calculation of $\Delta X_{i}$ and $\Delta X_{j}$.

For the interested reader, in Appendix C we illustrate how the solution of the diffusion equation may lead to the numerical loss of metals of up to 0.2\% in some extreme examples of isolated galaxies. This inaccuracy is caused by round off errors and the handling of particles using different time steps.

To calculate the diffusion coefficient, we adopt a coefficient that approximates the ``eddy diffusivity":

\begin{equation}\label{5}
D_{i} = C_{\rm{diffusion}}||S_{\alpha\beta}||_{i}\rho_{i}h_{i}^{2},
\end{equation}

\noindent where $C_{\rm{diffusion}}$ is a free parameter and $||S||$ denotes the Frobenius norm of the symmetric shear tensor. The coefficient $D_{i}$ was motivated by a Kolmogorov cascade by \citet{Smagorinsky63}, it depends on the velocity shear, hence it better models the mixing in shearing flows. 

The shear tensor is calculated as 

\begin{eqnarray}
S_{\alpha\beta}|_{i}&=&\frac{1}{2}(\tilde{S}_{\alpha\beta}|_{i}+\tilde{S}_{\beta\alpha}|_{i})-\delta_{\alpha\beta}\frac{1}{3}{\rm{Trace}({\bf{\tilde{S}}}}|_{i}),\\\label{6}
\tilde{S}_{\alpha\beta}|_{i}&=&\frac{1}{\rho_{i}}\sum_{j=1}^{N_{\rm{ngb}}}m_{j}(v_{\beta}|_{j}-v_{\beta}|_{i})\nabla_{\alpha,i}W(r_{ij},h_{i}),
\end{eqnarray}

\noindent where the sum is over SPH neighbours $j$, $W$ is the SPH kernel function, $v_{\beta}|_{j}$ is particle's velocity component in direction $\beta$, $\nabla_{i}$ is the gradient operator for particle $i$ (operating on the SPH kernel function) $\nabla_{\alpha,i}$ is the $\alpha$th component of the resultant vector, and $\delta_{\alpha\beta}$ is the Kronecker delta function.

$C_{\rm{diffusion}}$ needs to be calibrated for different numerical methods due to the different effective resolution scale of the turbulent cascade associated with each simulation. Typically, $C_{\rm{diffusion}}$ ranges from 0.1 to 0.2, as calculated from Kolmogorov theory (\citealt{Smagorinsky63, Wadsley08, Colbrook17}). In the FIRE simulations, $C_{\rm{diffusion}}$ was set to 0.15 (\citealt{Su17}), while in previous studies, it ranged from 0.05 to 0.1 (\citealt{Shen10,Brook14}). \citet{Escala18} used a value of $C_{\rm{diffusion}}=0.003$, calibrated to represent the minimum diffusion strength that results in significant sub-grid metal mixing and reproduces the width of the metallicity distribution function of local dwarfs. Increasing the diffusion strength above this value reduces the number of stars at both the high- and low-metallicity tails of the distribution. \citet{Sarrato23} used the Gasoline2 code in zoom-in cosmological simulations of MW-type galaxies and found that $C_{\rm{diffusion}}$ values between 0.003 and 0.01 best reproduce the metallicity dispersion observed in star clusters, H~\textsc{ii} regions, and neutral gas in the disc.

\citet{Rennehan19} introduced a different approach by developing a dynamic localized Smagorinsky model to simulate the turbulent diffusion of thermal energy and metal mixing. This method increased the metal-mixing timescale in the ISM. However, due to its computational cost, which requires iterations over neighboring particles within the kernel multiple times, we choose to follow the classical Smagorinsky model.

Finally, metal diffusion imposes a constraint on the time step, $\Delta t_{i}$, of a gas particle $i$ through the condition $\Delta t_{i}\leq 0.2h^{2}_{i}\rho_{i}/D_{i}$. This criterion is adopted to ensure the model reproduces the diffusion of a passive scalar in controlled test scenarios. An example is the diffusion of a sine-wave passive scalar field in a one-dimensional box with a uniform particle distribution of constant density and temperature, and no bulk motion.

In COLIBRE, the metal diffusion model is applied not only to the elemental mass fractions, but also to the mass fractions of dust grains, which are categorized by chemical composition and grain size, as well as to diagnostic tracers that monitor the mass contributions from various stellar mass loss channels, including AGB stars, massive stars and CCSNe, SNIa, and r-process enrichment sources. The fiducial value of $C_{\rm{diffusion}}$ is set to 0.01. In Section 4.2 we demonstrate how the value of $C_{\rm{diffusion}}$ affects the [Mg/Fe] abundance ratio of metal-poor stars in the Milky Way. 

\section{The COLIBRE Galaxy formation model}\label{Sec3}

\begin{table*}
\begin{center}
\caption{List of hydrodynamical simulations used in this study. From left-to-right the columns show: simulation name; comoving box size; number of initial gas particles particles (the number of dark matter particles is $4\times$N); mean dark matter particle mass; initial mean gas particle mass; Plummer-equivalent comoving gravitational softening length for; maximum proper softening length for gas; minimum supernova heating temperature; CCSN efficiency factor; normalization constant for the gas particle thermal pressure; metal diffusion coefficient. The subgrid parameters for energy feedback have been calibrated to reproduce the $z=0$ age galaxy stellar mass function and galaxy size-stellar mass relation.}
\begin{tabular}{ l|c|c|r|c|c|l|r|c|c|l|l }
\hline
Identifier & Box size & $N$ & $m_{\rm{DM}}$ & $m_{\rm{g}}$ & $\epsilon_{\rm{com}}$ & $\epsilon_{\rm{prop}}$  & $\Delta T_{\rm{SN,min}}$ & $f_{\rm{E,min}}$ & $P_{\rm{birth,0}}/k_{\rm{B}}$ & $C_{\rm{diffusion}}$ \\
& [Mpc] & & [$10^{6}$ M$_{\odot}$] & [$10^{6}$ M$_{\odot}$] & [kpc] & [kpc] & [$10^{6}$ K] & & [$10^{4}$ K cm$^{-3}$] & & \\
\hline
L025m7/Default diffusion & 25 & 188$^{3}$ & $19.40$ & $14.7$ & 3.58 & 1.4 & $3.16$ & 0.1 & 0.8 & 0.01 \\
L025m7/No diffusion & 25 & 188$^{3}$ & $19.40$ & $14.7$ & 3.58 & 1.4 & $3.16$ & 0.1 & 0.8 & 0.0 \\
L025m7/Low diffusion & 25 & 188$^{3}$ & $19.40$ & $14.7$ & 3.58 & 1.4 & $3.16$ & 0.1 & 0.8 & 0.001 \\
L025m7/High diffusion & 25 & 188$^{3}$ & $19.40$ & $14.7$ & 3.58 & 1.4 & $3.16$ & 0.1 & 0.8 & 0.1 \\
L025m6/Default diffusion & 25 & 376$^{3}$ & $2.42$ & $1.84$ & 1.79 & 0.7 & $5.62$ & 0.3 & 1.0 & 0.01 \\
L025m6/No diffusion & 25 & 376$^{3}$ & $2.42$ & $1.84$ & 1.79 & 0.7 &  $5.62$ & 0.3 & 1.0 & 0.0 \\
L025m6/Low diffusion & 25 & 376$^{3}$ & $2.42$ & $1.84$ & 1.79 & 0.7 &  $5.62$ & 0.3 & 1.0 & 0.001 \\
L025m6/High diffusion & 25 & 376$^{3}$ & $2.42$ & $1.84$ & 1.79 & 0.7 &  $5.62$ & 0.3 & 1.0 & 0.1 \\
L050m6/Default diffusion & 50 & 752$^{3}$ & $2.42$ & $1.84$ & 1.79 & 0.7 &  $5.62$ & 0.3 & 1.0 & 0.01 \\
L012m5/Default diffusion & 12.5 & 376$^{3}$ & $0.30$ & $0.23$ & 0.89 & 0.35 &  $10.00$ & 1.0 & 3.0 & 0.01 \\
\hline
\end{tabular}
\end{center}
\label{Hydro_simulations}
\end{table*}

In this work, we analyse a series of cosmological simulations with periodic comoving volumes of (12.5 Mpc)$^{3}$, (25 Mpc)$^{3}$ and (50 Mpc)$^{3}$, using the open-source astrophysical code SWIFT\footnote{\href{www.swiftsim.com}{www.swiftsim.com}} (\citealt{Schaller23}). The SWIFT gravity solver employs the Fast Multipole Method (\citealt{Greengard87}) with an adaptive opening angle for short-range forces and a particle-mesh method solved in Fourier space for long-range forces. For hydrodynamics, SWIFT employs the energy-density SPHENIX SPH scheme (\citealt{Borrow22}). 

The simulations follow the evolution of $N$ gas particles and $4\times N$ dark matter particles up to redshift $z=0$. In our fiducial model $N=376^3$ over a (25 Mpc)$^{3}$ comoving volume. For this model, the comoving softening for gas, $\epsilon_{\rm{com}}$, is set to 1.79 kpc (2.09 kpc for dark matter) at early times, and limited to a physical value of $\epsilon_{\rm{prop}}=700$ pc (819 pc for dark matter) at $z\leq 1.56$. The dark matter particle mass is $2.42\times 10^{6}~\rm{M}_{\odot}$, and the initial gas particle mass is $1.84\times 10^{6}~\rm{M}_{\odot}$. See Table~\ref{Hydro_simulations} for variations of these values depending on resolution. The starting redshift of the simulations is $z=127$. The initial conditions were generated using MONOFONIC (\citealt{Hahn20}), see \citet{Schaye25} for details. The adopted cosmological parameters are $\Omega_{\rm{m}}=0.306$, $\Omega_{\Lambda}=0.694$, $\Omega_{\rm{b}}=0.0486$, $\Omega_{\nu}=0.00139$, $X=0.756$, $h=0.681$, $\sigma_{8}=0.807$ and $n_{\rm{s}}=0.967$ (\citealt{Abbott22}). We assume a single massive neutrino species with a mass of 0.06 eV. The COLIBRE model adopts a small minimum SPH smoothing length, $h_{\rm{min}}$, of $h_{\rm{min}}= 10^{-8}\epsilon_{\rm{soft}}$, to avoid potential run-away collapse of cold, dense gas (see \citealt{Ploeckinger24}).

Note that the COLIBRE simulation suite contains larger cosmological boxes both in terms of size and resolution (see \citealt{Schaye25}). However, due to the computational cost associated with running large-volume, high-resolution simulations, we conducted our analysis using smaller box sizes and lower resolution. This approach enabled us to systematically explore the impact of varying parameters such as the diffusion coefficient, which would be prohibitively expensive to investigate in larger-scale simulations.

The simulations incorporate various physical processes, including radiative gas cooling and heating, dust, cosmic rays and self-shielding prescriptions from \citet{Ploeckinger20}, including the following updates on the self-shielding prescriptions of the cold, dense gas. Radiative cooling and heating processes are modelled with hybrid-chimes (\citealt{Ploeckinger25}), which we use to follow the non-equilibrium species abundances of electrons and nine hydrogen and helium species. The cooling rates from the remaining 147 metal species are taken from pre-computed tables that were calculated using the full CHIMES network assuming chemical equilibrium, scaled by the non-equilibrium electron abundance. The engine for the calculations of the cooling and heating rates is the chemical network solver CHIMES (\citealt{Richings14a,Richings14b}), which is integrated in SWIFT for the non-equilibrium chemistry calculations, and also used as a stand-alone software package to calculate the pre-tabulated equilibrium properties of the metal species. The redshift-dependent UV/X-ray background as well as the parametrisations of the interstellar radiation field (ISRF), the cosmic ray rate, and the shielding column density are based on those used in \citet{Ploeckinger20}, with the updates described in (\citealt{Ploeckinger25}). Note that the fiducial \citet{Ploeckinger25} normalisation of the ISRF, which we adopt here, is a factor 10 larger than the value used in the fiducial COLIBRE simulations presented in \citet{Schaye25}.


The COLIBRE model incorporates a subgrid model for the formation and evolution of interstellar dust (\citealt{Trayford25}). The model includes the evolution of dust grains: graphites and silicates, produced in the AGB phase of stellar evolution and in CCSNe. Dust grains evolve by accreting gas, by grain shattering and coagulation, by thermal sputtering, and can be destroyed due to shocks from SNe. We assume that all dust grains have spherical shapes and track two sizes, 0.01 and 0.1 $\mu$m. In addition, the model includes the effects of metal depletion. The dust obeys an overall metal budget, both in terms of stellar yields and within a given gas element, consistent with the model presented in this work.

Gas is allowed to cool down to a minimum temperature of $\approx$10 K. To be eligible for star formation, gas must satisfy the gravitational instability criterion following \citet{Nobels23}. The star formation rates of eligible gas particles are based on the \citet{Schmidt59} law, $\dot{\rho}_{*}=\epsilon\frac{\rho}{t_{\rm{ff}}}$, where $\epsilon=0.01$ is the star formation efficiency per free-fall time, $t_{\rm{ff}}=\sqrt{3\pi/(32G\rho)}$. Once formed, star particles produce various forms of stellar feedback, such as stellar winds, H~\textsc{ii} regions and SNe. To determine the energies, momenta, and ionizing flux injected into the gas phase by pre-SN processes, we use the Binary Population and Spectral Synthesis (BPASS) tables (\citealt{Eldridge17,Stanway18}) version 2.2.1 with a \citet{Chabrier03} IMF (see \citealt{Benitez25}). 

Stellar feedback from CCSNe follows a modified version of the thermal-kinetic model of \citet{Chaikin23}. Stars with ZAMS masses greater than 8 $\rm{M}_{\odot}$ explode as CCSN, releasing $\Delta E_{\rm{CCSN}}=f_{\rm{E}}\times 10^{51}$ ergs of energy, where $f_{\rm{E}}$ depends on the thermal pressure of the parent gas particle, $P_{\rm{birth}}$, as well as on the minimum and maximum efficiency factors, $f_{\rm{E,min}}$ and $f_{\rm{E,max}}$, respectively, as follows

\begin{equation}
f_{\rm{E}}=f_{\rm{E,min}}+\frac{f_{\rm{E,max}}-f_{\rm{E,min}}}{1+{\rm{exp}}[-\log_{10}(P_{\rm{birth}}/P_{\rm{birth,0}})/\sigma_{\rm{P}}]}.
\end{equation}

\noindent Here, $P_{\rm{birth,0}}$ is a normalization constant and $\sigma_{\rm{P}}$ is a parameter that defines the width of the transition function from $f_{\rm{E,min}}$ to $f_{\rm{E,max}}$. The functional form of $f_{\rm{E}}$ indicates that SN feedback of stellar particles born in higher pressure environments will be more energetic. The values of $f_{\rm{E,min}}$ and $P_{\rm{birth,0}}$ vary with resolution (see Table~\ref{Hydro_simulations} for details), while $f_{\rm{E,max}}$ is set to 4 and $\sigma_{\rm{P}}$ to 0.3. These parameters have been calibrated to reproduce the $z=0$ age galaxy stellar mass function and galaxy sizes. For further details on the calibration process of the simulations, refer to \citet{Chaikin25}. For the complete list of parameters, also refer to table 1 in \citet{Schaye25}. Note that the m5-resolution simulations produced for this study differ from the fiducial simulations presented in \citet{Schaye25}. In the fiducial COLIBRE model at
m5 resolution, the parameters are set to $f_{\rm{min}} = 0.8$ and $P_{\rm{birth,0}}/k_{\rm{B}} = 1.5\times 10^{4}$ K cm$^{-3}$, whereas this study adopts slightly different values (indicated in Table 3). This difference arises because our m5 simulation was generated before the feedback calibration of the fiducial model had been completed.

The $\Delta E_{\rm{CCSN}}$ energy is transferred stochastically to the surrounding gas within the SPH kernel of the stellar particle using the isotropic scheme of \citet{Chaikin22}. As in \citet{Chaikin23}, an additional parameter $f_{\rm{kin}}=0.1$, is used to split $\Delta E_{\rm{CCSN}}$ into two channels of energy injection: thermal and kinetic. A fraction $f_{\rm{kin}}\Delta E_{\rm{CCSN}}$ is injected kinetically, while $(1-f_{\rm{kin}})\Delta E_{\rm{CCSN}}$ is injected in thermal form. The thermal channel of CCSN feedback follows the stochastic model of \citet{DallaVecchia12}, including the isotropic energy injection from \citet{Chaikin22}. Note, however, that the COLIBRE model does not use a constant heating temperature, instead it allows the heating temperature to vary within a range of values, $\Delta T_{\rm{SN,min}}<\Delta T_{\rm{SN}}<\Delta T_{\rm{SN,max}}$, that monotonically increases with gas density. In our fiducial model, we set $ \Delta T_{\rm{SN,min}}=10^{6.75}$ K (which varies with resolution) and $\Delta T_{\rm{SN,max}}=10^{8}$ K. In the kinetic channel of energy injection, stellar particles stochastically kick neighbouring gas particles with a target velocity of 50 km s$^{-1}$ in a manner that explicitly conserves energy, linear momentum, and angular momentum (see \citealt{Chaikin23} for details). Stellar feedback produced by SNIa events follows the implementation of CCSN feedback but with $f_{\rm{kin}}=0$ and $f_{\rm{E}}=1$.

The final ingredient in the COLIBRE model are supermassive black holes. The formation and evolution of black holes are modelled following the approach of \citet{Springel05b, Booth09, Bahe22}, though with some modifications. Initially seeded within friends-of-friends dark matter groups, black holes' gas accretion rates follows the \citet{Krumholz05} modified Bondi-Hoyle-Lyttleton formation for accretion rate that accounts for supersonic turbulence and vorticity. The feedback mechanism from active galactic nucleus (AGN) activity is implemented thermally following \citet{Booth09}. The energy depends on the accreted mass, $\dot{m}_{\rm{BH}}$, onto the black hole, and it is stored in a reservoir carried by each black hole particle until it can be utilized to heat the nearest neighbouring gas particle, inducing a temperature increase of $\Delta T_{\rm{AGN}}$ that scales with black hole mass, $\Delta T_{\rm{AGN}}=10^{9}~{\rm{K}}~(m_{\rm{BH}}/10^{8}\rm{M}_{\odot})$ (with $\Delta T_{\rm{AGN}}$ having minimum and maximum values of $3.16\times 10^{6}$ K and $10^{10}$ K, respectively). 

Table~\ref{Hydro_simulations} provides an overview of the values of the parameters used in the different simulations in this work, along with their name identifiers. For further details the reader is referred to \citet{Schaye25}. 

Halo catalogues were generated using the VELOCIraptor halo finder (\citealt{Elahi11,Elahi19,Canas19}). VELOCIraptor uses a 3D-friends of friends (FOF) algorithm to identify field haloes, and subsequently applies a 6D-FOF algorithm to separate virialised structures and identify sub-haloes of the parent haloes (\citealt{Elahi19}).

\section{Calibration of the Chemical enrichment model}\label{Sec4}

The free parameters that control the chemical enrichment and the metal distribution in the COLIBRE model are the boost factors for the CCSN yields, the metal diffusion coefficient, as well as the DTD function of SNIa explosions. For the SNIa free parameters, we follow the work of Nobels et al. (in prep.), who investigated their impact on the stellar abundance relations and determined the fiducial values assumed in this work (see Section 2.1.5 for more details). In this section, we focus on analysing the impact of the boost factors for the CCSN yields and the metal diffusion coefficient.

\subsection{CCSN boost factors}\label{CCSN_boost_factors_section}

\begin{figure*} 
\begin{center}
\includegraphics[angle=0,width=\textwidth]{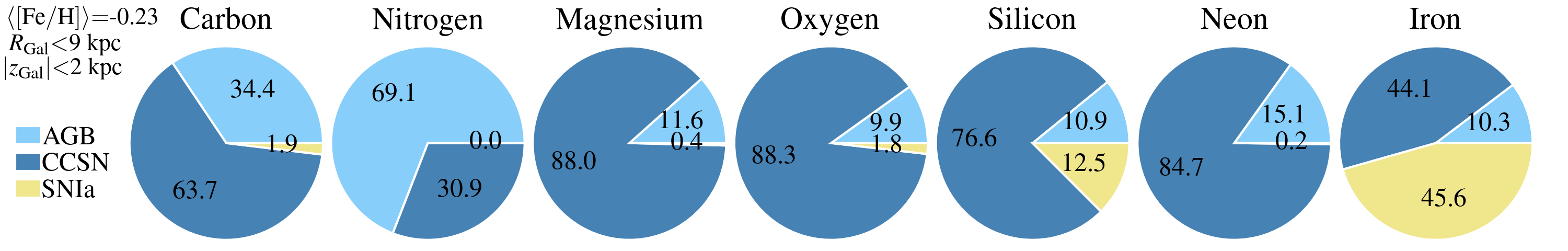}\\
\includegraphics[angle=0,width=\textwidth]{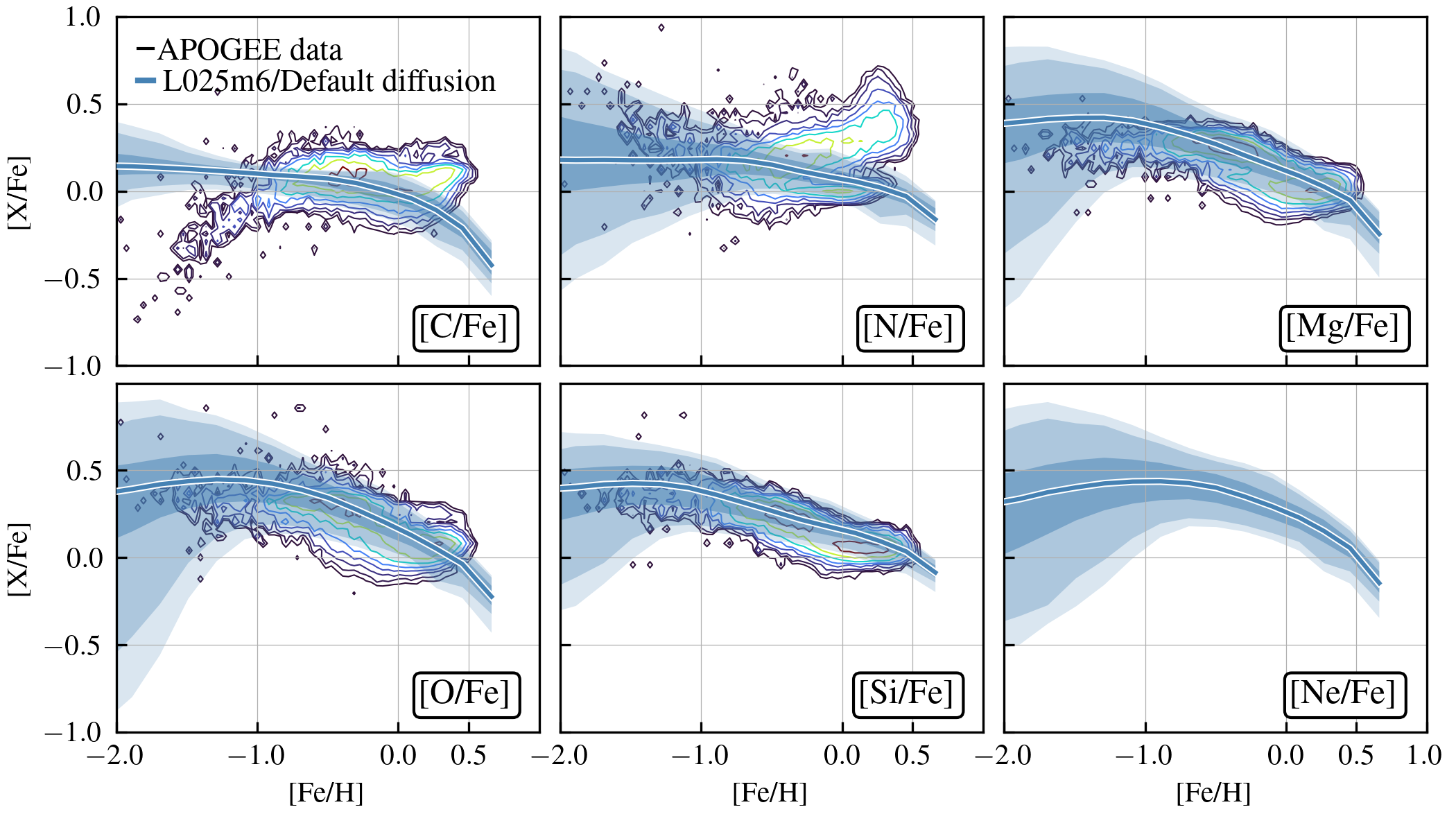}\\
\caption{Element abundance ratios from simulated star particles with Galactocentric radii smaller than 9 kpc and azimutal distances under 2 kpc. The element abundances of these star particles, characterized by a median metallicity of [Fe/H]=-0.23, are compared to observed abundances in the Milky Way. {\it{Top panels:}} Pie charts that show the percentage of each element that was produced by AGB stellar enrichment (light blue), CCSN (dark blue), and SNIa (yellow). {\it{Middle and bottom panels:}} element abundance ratios for Milky Way-mass galaxies from the simulation (L025m6/Default diffusion). Each panel corresponds to a different element: carbon, nitrogen, and magnesium (middle panels), as well as oxygen, silicon, and neon (bottom panels), all plotted as a function of iron over hydrogen. Solid lines represent the simulated median relations, while shaded regions mark the 16-84th, 5-95th, and 1-99th percentiles. The contour plots show weighted Milky Way stellar abundances from the APOGEE DR14 data release.}
\label{StellarAbundance}
\end{center}
\end{figure*}

Previous studies (e.g., \citealt{Wiersma09, Philcox18, Buck21,Liang23}) have shown that chemical yields vary significantly between different studies, even for a fixed IMF and without accounting for stellar rotation or binary evolution. Given this uncertainty, we decided to adjust the stellar yields to better reproduce observed $z=0$ stellar abundance ratios. Our observational targets are the abundance ratios of stars in the Galactic disk, based on data from the APOGEE survey (\citealt{Majewski17}).

APOGEE is a large-scale near-infrared stellar spectroscopic survey. We use data release DR14 (\citealt{Jonsson18}) and cross-match it with the astroNN added-value catalog (\citealt{Leung19}) to obtain individual star Galactocentric positions from Gaia eDR3 (\citealt{GaiaCollaboration21}), which were calculated assuming the Sun is located 8.125 kpc from the Galactic center (\citealt{GRAVITYCollaboration18}) and 20.8 pc above the Galactic midplane (\citealt{Bennett19}). Since APOGEE primarily samples stars near the solar neighborhood, we minimize this bias by adjusting the observed stellar distribution. We create six distributions based on Galactocentric radial position, $R_{\rm{Gal}}$, and vertical height, $z_{\rm{Gal}}$: $R_{\rm{Gal}}=0{-}3$, $3{-}6$, and $6{-}9$ kpc, and $|z_{\rm{Gal}}|=0{-}1$ and $1{-}2$ kpc. We add these to derive a joint stellar abundance distribution that gives less weight to stars in the solar vicinity. This is because each bin in $z_{\rm{Gal}}$ and $R_{\rm{Gal}}$ contributes equally to the final distribution, regardless of how many stars are in each bin. The resulting contours, shown in the middle and bottom panels of Fig.~\ref{StellarAbundance}, use a logarithmic scale with a bin size of 0.2. For neon, observational data is not available due to the difficulty of obtaining reliable measurements.

Fig.~\ref{StellarAbundance} compares simulated element abundance ratios from the L025m6/Default diffusion model with observed stellar abundances in the Milky Way. The top panels show the fractional mass from each element produced by different stellar enrichment processes: AGB (light blue), CCSN (dark blue) and SNIa (yellow). The middle and bottom panels show the stellar abundance ratios, [X/Fe]\footnote{We remind that the notation [X/Fe] indicates the logarithm base 10 of the abundance ratio of element X to iron relative to solar values.}, as a function of iron over hydrogen ([Fe/H]), for carbon, nitrogen, magnesium, oxygen, silicon and neon, as indicated in each panel. The solid and shaded regions show the median and percentile distributions of stellar abundance ratios from Milky Way-mass galaxies in the simulation, while the contours represent APOGEE stellar abundances. 

For a fair comparison with APOGEE data, we select 27 Milky Way-like galaxies from the L025m6/Default diffusion model based on their halo mass ($[0.6-1]\times 10^{12}~\rm{M}_{\odot}$), stellar mass ($[1-6]\times10^{10}~\rm{M}{\odot}$), and disc morphology. Stellar mass is defined by summing up all star particles within a 3D aperture with radius of 50 kpc. Halo mass is defined as the total mass of all particles enclosed within the radius at which the mean internal density is 200 times the critical density of the Universe. Morphology is determined based on the kinematic indicator presented in \citet{Correa17}. From these galaxies, we select all star particles at $R_{\rm{Gal}}{<}9$ kpc and $|z_{\rm{Gal}}|{<}2$ kpc, and examine their metal composition.

To better reproduce the median APOGEE abundances in the metallicity range $-1{\leq}[\rm{Fe}/\rm{H}]{\leq}1$, we adjusted the CCSN yields with boost factors: 1.5 for carbon and magnesium, and 1 (no boost) for the other elements. This adjustment allows the model to closely reproduce the observed element ratios, especially for $\alpha$-elements. However, the model fails to capture certain key features observed in the element abundances. For carbon, the observed increasing [C/Fe] ratio for $[\rm{Fe}/\rm{H}]{>}0$ is absent in the simulations. In the observations, nitrogen shows a complex bimodal distribution: one trend where [N/Fe] increases with [Fe/H] and another with a constant $[\rm{N}/\rm{Fe}]{=}0$. These intriguing features are also missing in our model. APOGEE shows that $\alpha$-elements exhibit a bimodal distribution (e.g. \citealt{Adibekyan12, Hayden15, Vincenzo21}), commonly attributed to an inner-disc with low [$\alpha$/Fe] and young stellar population, and an outer-disc with higher [$\alpha$/Fe] and older stellar population (e.g. \citealt{BlandHawthorn19, Sharma21, Imig23}). Our models correctly follow the median trend of decreasing [$\alpha$/Fe] with increasing [Fe/H], with [$\alpha$/Fe] being traced by [Mg/Fe]. We leave for the future a detailed analysis of a potential bimodal distribution, which previous work suggests reflects the Milky Way's gas accretion history (e.g. \citealt{Mackereth18}).


Next, we examine the contributions that AGB stars, CCSNe, and SNIa make to the total metal mass in Milky Way stars. To account for the variation within the Milky Way sample, we calculate the fractional mass of each element produced by AGB, CCSNe, and SNIa in each galaxy, and then we take the average across the sample. The top row of Fig.~\ref{StellarAbundance} shows that for an average stellar metallicity of $\langle[\rm{Fe}/\rm{H}]\rangle = -0.23$, AGB stars contribute approximately 34\% of the carbon and 69\% of the nitrogen, with the remainder primarily produced by CCSNe and for carbon a small contribution of about $2\%$ from SNIa. These values are lower than those predicted by the Galactic chemical evolution model of \citet{Kobayashi20} (see also \citealt{Kobayashi23}), which estimates that 49\% of the carbon and 74\% of the nitrogen in the solar neighborhood are produced by AGB stars (at $t = 9.2$ Gyr). Although their model focuses on the solar vicinity, our analysis incorporates a broader range of Galactic environments. The $\alpha$-elements (Mg, O, Si, and Ne) are predominantly produced by CCSNe, consistent with the findings of \citet{Kobayashi20}.

Iron is distributed more evenly over the enrichment channels: 46\% from SNIa, 44\% from CCSN, and a smaller, non-negligible fraction (${\approx}10\%$) from AGB stars. For comparison, \citet{Kobayashi20} predict that 60\% of iron is produced by SNIa, with the rest from CCSN. When we change the selection criteria to stars at $R_{\rm{Gal}}{>}9$ kpc and $|z_{\rm{Gal}}|{>}2$ kpc (halo stars), the average metallicity decreases to $\langle[\rm{Fe}/\rm{H}]\rangle = -1.47$. While the relative contributions from AGB stars, CCSN, and SNIa remain similar to those for disc stars, iron shows a slight change: 44\% is produced by SNIa, 49.7\% by CCSN, and about 6.3\% by AGB stars.

\subsubsection{$\alpha$-enhancement}

CCSNe primarily release $\alpha$-elements over short timescales (e.g. \citealt{Woosley95}). In our models, CCSNe contribute to 76\%, 85\% and 88\% of the total mass of Si, Ne and O and Mg, respectively, as shown in Fig.~\ref{StellarAbundance}. Iron is more evenly distributed between CCSNe and SNIa, with SNIa producing it over longer timescales compared to CCSNe. As a result, the $\alpha$-element-to-iron abundance ratios (also referred to as $\alpha$-enhancement when compared to solar abundance ratios) remain roughly constant at low metallicities until SNIa enrichment becomes significant. In our models, the average Fe mass fraction from SNIa exceeds 0.5 for star particles with $[\rm{Fe}/\rm{H}]{\geq}0$. 

\begin{figure} 
\begin{center}
\includegraphics[angle=0,width=0.45\textwidth]{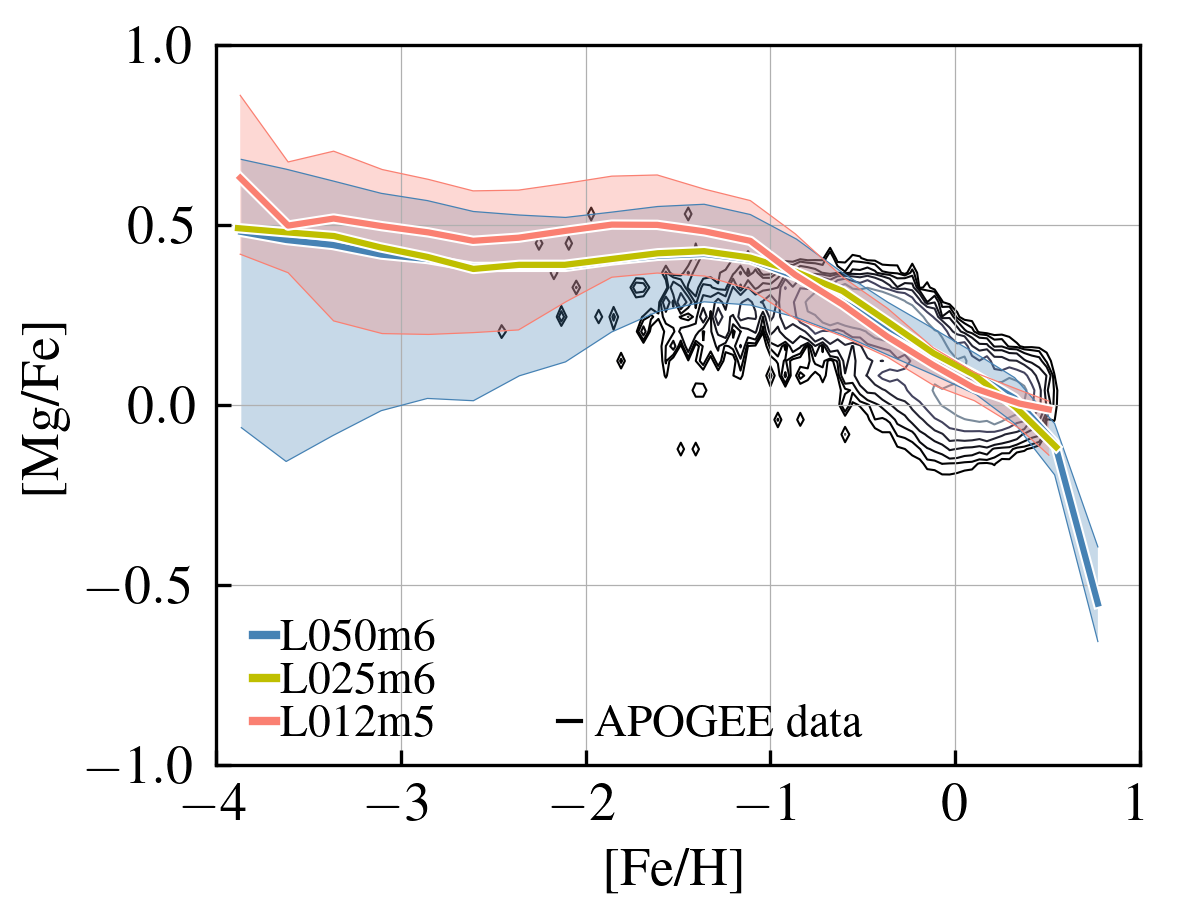}
\includegraphics[angle=0,width=0.45\textwidth]{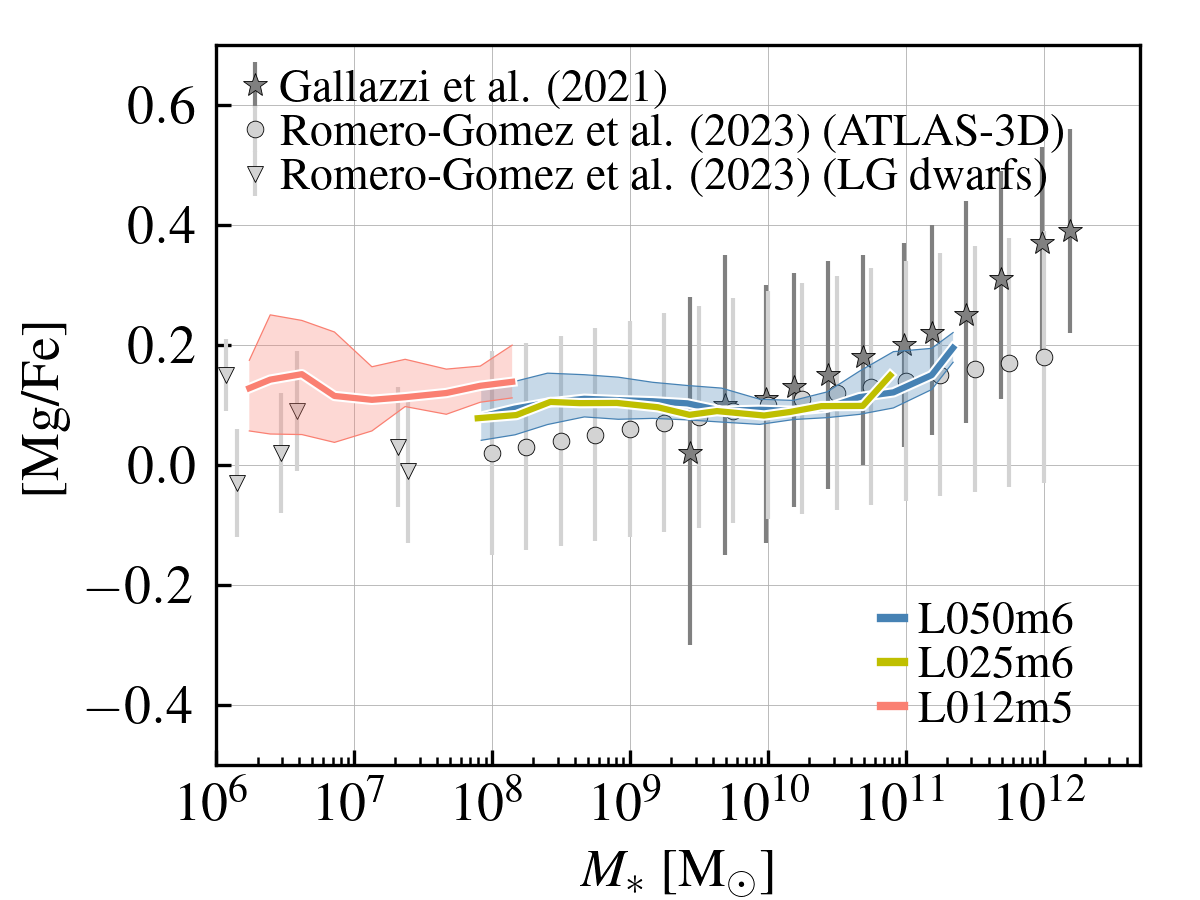}
\caption{$\alpha$-enhancement relations. {\it{Top panel}}: [Mg/Fe] vs. [Fe/H] relation for all stars within Milky Way-like galaxies in the COLIBRE simulations with different box sizes and/or resolutions, as indicated in the legend. The solid colored lines represent the median relation, with shaded regions showing the 16th to 84th percentiles. Black contours correspond to the weighted stellar distribution from the APOGEE survey. {\it{Bottom panel}:} Total stellar [Mg/Fe] from centrals galaxies as a function of the galaxies' stellar mass. As in the top panel, solid lines show the median relations from different cosmological simulations. Grey symbols correspond to observations from \citet{Gallazzi21} (stars symbols), and \citet{Romero23}, including data from massive early-type galaxies from the ATLAS-3D survey (circles) and dwarf galaxies from the Local Group (triangles).}
\label{alphaenhancement}
\end{center}
\end{figure}

The top panel of Fig.~\ref{alphaenhancement} shows the stellar [Mg/Fe] ratio as a function of [Fe/H] for Milky Way-mass galaxies from the COLIBRE simulations with different box sizes and resolutions (see details in Section 3). The black contours correspond to the weighted APOGEE stellar abundances as shown in Fig.~\ref{StellarAbundance}. In the L025m6/Default diffusion model, $[\rm{Mg}/\rm{Fe}]\approx 0.4$ for $[\rm{Fe}/\rm{H}]{<}-1$. As metallicity increases, iron from SNIa becomes more significant, and [Mg/Fe] decreases, reaching $[\rm{Mg}/\rm{Fe}]=0.1$ at $[\rm{Fe}/\rm{H}]=0$, and $[\rm{Mg}/\rm{Fe}]=0$ at $[\rm{Fe}/\rm{H}]=0.2$. 

The panel shows good convergence with the simulation volume between the L025m6 and L050m6 models, which share the same numerical resolution. When resolution is increased, numerical convergence is roughly maintained within the metallicity range $-1{\leq}[\rm{Fe}/\rm{H}]{\leq}1$. However, for $[\rm{Fe}/\rm{H}]<{-}1.5$, the higher-resolution models show $[\rm{Mg}/\rm{Fe}]{\approx}0.5$, indicating a higher constant value for $\alpha$-enhancement. At $[\rm{Fe}/\rm{H}]=0$, $[\rm{Mg}/\rm{Fe}]=0.11$ from L025m6 and $[\rm{Mg}/\rm{Fe}]=0.08$ from the the higher resolution L012m5 simulation.

To determine the boost factor for Mg, we analyzed not only the abundance ratios of stars in Milky Way-like galaxies, but also the total $\alpha$-enhancement of galaxies as a function of stellar mass. The bottom panel of Fig.~\ref{alphaenhancement} shows the total stellar [Mg/Fe] of $z=0$ central galaxies from various cosmological models. We compute [Mg/Fe] by summing the total magnesium and iron content from the stellar component of each galaxy, then taking the logarithm of their abundance ratio relative to solar values. We compare the median relations with observational data from \citet{Gallazzi21} and \citet{Romero23}. \citet{Gallazzi21} combined the SDSS group catalogue with stellar populations parameters estimated for SDSS DR7 galaxies, following \citet{Gallazzi05}. They empirically estimated [$\alpha$/Fe] by measuring the index ratio Mgb/$\langle\rm{Fe}\rangle$ on the observed spectrum and comparing it with the index ratio from model libraries. The excess $\Delta$(Mgb/$\langle\rm{Fe}\rangle$) in the data relative to the models is interpreted as an excess of [$\alpha$/Fe] with respect to solar. \citet{Romero23} analyzed the spectra of each ATLAS-3D galaxy using the MILES/Vazdekis stellar templates library. For the dwarfs, they presented a compilation of stellar abundances taken from \citet{Simon19} and \citet{McConnachie20}. For this study, all observational data were converted to the solar abundances of \citet{Asplund09}.

We find that a boost factor of 1.5 for magnesium allows us to reproduce both, the extragalactic [Mg/Fe] and the individual stellar [Mg/Fe] abundances from Milky Way stars. Specifically, our models match the total [Mg/Fe] for galaxies with a stellar mass ${\sim}10^{10}~\rm{M}_{\odot}$, as seen in the comparison with the observational data from Fig.~\ref{alphaenhancement}. For higher-mass galaxies, we expect [Mg/Fe] to increase with stellar mass because massive galaxies form a large majority of their stars over a period of time similar to or shorter than the SNIa delay time due to quenching of their star formation rates (see e.g. \citealt{Segers16}). However, due to the limited box size of this study, we lack a robust sample of high-mass galaxies to fully test this effect. For galaxies with stellar masses below $10^{10}~\rm{M}_{\odot}$, we observe an increase in [Mg/Fe] for dwarf galaxies, potentially forming a U-shaped trend, which has been reported by \citet{Romero23}. Further analysis of the environment surrounding these dwarf galaxies and central/satellite differences is needed to determine and interpret the behavior of [Mg/Fe] at low masses.

\subsection{Diffusion coefficient}\label{Sec42}

\begin{figure*} 
\begin{center}
\includegraphics[angle=0,width=0.9\textwidth]{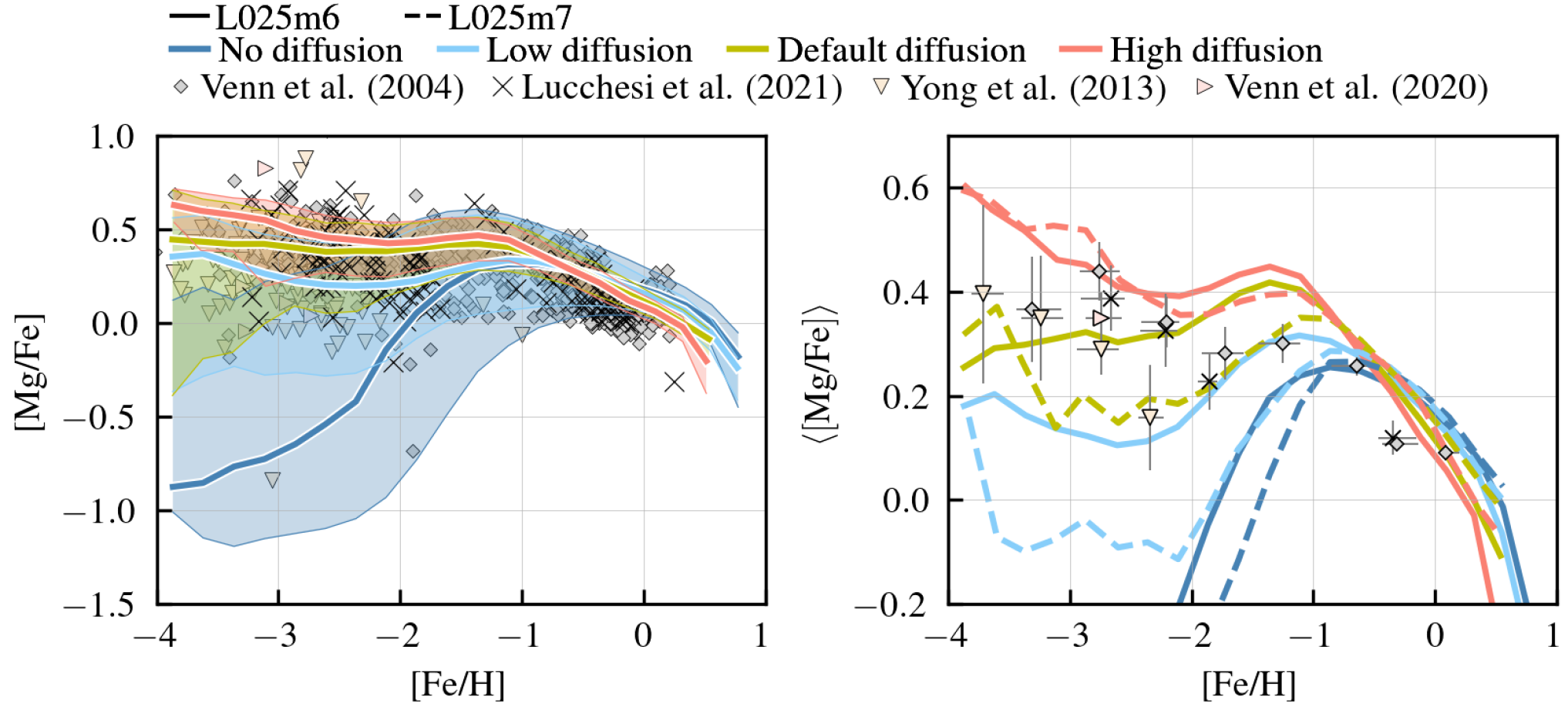}
\caption{Analysis of metal-poor stars in the Milky Way. {\it{Left panel}}: [Mg/Fe] vs. [Fe/H] for star particles in Milky Way-like galaxies. Solid lines represent the median values, and the shaded regions show the 16th to 84th percentiles from simulations with different diffusion coefficients: high ($C_{\rm{diffusion}}=0.1$, orange), default ($C_{\rm{diffusion}}=0.01$, light green), low ($C_{\rm{diffusion}}=0.001$, light blue), and no diffusion ($C_{\rm{diffusion}}=0$, dark blue). The diffusion treatment significantly impacts the abundance patterns of metal-poor stars ($[\rm{Fe}/\rm{H}]<{-}1$). Observational data from the Milky Way are taken from \citet{Venn04}, \citet{Yong13}, and the Pristine Survey (\citealt{Venn20, Lucchesi22}). {\it{Right panel}}: Comparison to the average [Mg/Fe] as a function of [Fe/H] for Milky Way stars. Note that the axis range differs from the left panel. The observational data points show the mean values in different metallicity bins, with error bars corresponding to the 95\% confidence intervals. The coloured lines indicate the mean [Mg/Fe] of star particles from simulations. Solid lines correspond to L025m6 simulations, and dashed lines to L025m7 simulations. Metal diffusion mostly impacts the chemical abundances of metal-poor stars.}
\label{DiffusionAnalysisPlot}
\end{center}
\end{figure*}

In this section, we explore how our treatment of metal mixing affects the variation in element abundances among star particles. We again focus on the abundance ratio [Mg/Fe] as a function of [Fe/H] in star particles from Milky Way-like galaxies, serving as a proxy for the overall enrichment of $\alpha$-elements. We select galaxies with halos of mass $[0.6-1]{\times}10^{12}~\rm{M}_{\odot}$ without applying any further cuts on stellar mass or morphology, as these factors have minimal impact on the results. We then select all star particles gravitationally bound to the central subhalo, i.e. excluding those bound to satellite galaxies. Additionally, no selection is made based on $R_{\rm{Gal}}$ or $z_{\rm{Gal}}$ (as in Section~\ref{CCSN_boost_factors_section}), since we aim to include both disc star particles and metal-poor halo stars.

The left panel of Fig.~\ref{DiffusionAnalysisPlot} shows the resulting [Mg/Fe] as a function of [Fe/H]. The solid lines represent the median [Mg/Fe]-[Fe/H] relations from the L025m6 simulations for different diffusion coefficients: high ($C_{\rm{diffusion}}{=}0.1$, orange), default ($C_{\rm{diffusion}}{=}0.01$, light green), low ($C_{\rm{diffusion}}{=}0.001$, light blue), and no diffusion ($C_{\rm{diffusion}}{=}0$, dark blue). Across the metallicity range $-1{\leq}\rm{[Fe/H]}{\leq}1$, metal diffusion has only a small impact on the median [Mg/Fe] ratio. However, at lower metallicities ($\rm{[Fe/H]}{<}-1$), diffusion effects become more significant. In the L025m6/No diffusion model, many metal-poor star particles ($\rm{[Fe/H]}{<}-1$) have [Mg/Fe] values between $-1.0$ and $0.5$, while the L025m6/High diffusion model predicts values closer to $0.5$. This demonstrates that metal diffusion primarily affects the chemical abundances of metal-poor stars, which can be understood if, as expected, lower metallicity particles have experienced a smaller number of enrichment events.

The left panel of Fig.~\ref{DiffusionAnalysisPlot} also includes observational data from Milky Way stars. These are sourced from \citet{Venn04} (grey diamonds), from \citet{Yong13} (light orange triangles), and from the Pristine Survey by \citet{Venn20} (light pink triangles) and \citet{Lucchesi22} (black crosses). \citet{Venn04} compiled Galactic stellar abundance data from multiple studies (\citealt{Edvardsson93,McWilliam95,Ryan96,Nissen97,Hanson98,Fulbright00,Prochaska00,Stephens02,Reddy03,Bensby03}) alongside kinematic data, showing that most metal-poor stars in their sample reside in the Galactic halo. \citet{Yong13} analyzed the chemical abundances of 190 metal-poor Galactic halo stars, including 171 stars with [Fe/H]${<}-2.5$, of which 86 are extremely metal-poor ([Fe/H]${<}-3$). The Pristine survey is a photometric survey designed to efficiently pre-select very metal-poor stars. \citet{Venn20} carried out a first chemo-dynamical analysis of 115 metal-poor stars from this survey using Bayesian inference to determine chemical abundances. \citet{Lucchesi22} improved this work with better data reduction and metallicity calibration. \citet{Lucchesi22} and \citet{Yong13} assumed \citet{Asplund09} solar abundances, while \citet{Venn20} adopted MESA/MIST stellar isochrones with solar-scaled compositions from \citet{Grevesse93}. To ensure consistency, we converted the stellar abundances from \citet{Venn20} to \citet{Asplund09} solar values. The compilation by \citet{Venn04} predominantly assumes solar abundances from \citet{Anders89} or \citet{Grevesse96}, which we also converted accordingly.

As described in Section~\ref{Metal_diffusion_Sec}, our model uses the metal diffusion coefficient, $C_{\rm{diffusion}}$, as a free parameter to control the rate of metal mixing due to unresolved turbulent diffusion. Previous studies (e.g., \citealt{Escala18, Sarrato23}) have constrained metal mixing by comparing to the observed star-to-star scatter in abundance ratios. However, this approach has a key limitation. Our model represents stellar populations under a given IMF with macro-particles, rather than individual stars. This means that when estimating element abundances for star particles, we are effectively averaging over many stars, with a sampling noise determined by the resolution scale. Therefore, the scatter between star particles in our model cannot be directly compared to the observed star-to-star scatter. To constrain $C_{\rm{diffusion}}$, we instead focus on matching the average [Mg/Fe] of metal-poor stars. The mean is particularly sensitive to the overall scatter and presence of outliers in the sample, and $C_{\rm{diffusion}}$ has the largest impact on the mean [Mg/Fe] for stars with [Fe/H]${<}{-}1$.

The right panel of Fig.~\ref{DiffusionAnalysisPlot} shows the average [Mg/Fe] as a function of [Fe/H]. The different coloured solid lines represent L025m6 simulations with different diffusion coefficients, while the dashed lines now correspond to lower-resolution L025m7 simulations. For the observational data, we binned stars in [Fe/H] intervals of 0.5 dex, selected bins with at least 10 stars, and calculated the mean [Mg/Fe] and 95\% confidence intervals (${\pm}1.96~\sigma/\sqrt{N}$), for which we assumed that the underlying population distribution of [Mg/Fe] is normal.

In the metallicity range $-2{\leq}\rm{[Fe/H]}{\leq}{-}1$, the L025m6/Low diffusion and L025m7/Default diffusion models agree well with the observations, while the L025m6/Default diffusion model overshoots the data by 0.1 dex. At lower metallicities ($-4{\leq}\rm{[Fe/H]}{\leq}{-}2$), the observed mean [Mg/Fe] rises to ${\approx}0.35$, bringing the L025m6/Default diffusion model into better agreement with the data. The L025m6/High diffusion model predicts excessively high [Mg/Fe] values, except in the range $-3{\leq}\rm{[Fe/H]}{\leq}{-}2$, while the L025m6/Low diffusion model produces a [Mg/Fe] mean that is low relative to observations.

Interestingly, in the range $-1{\leq}\rm{[Fe/H]}{\leq}1$, the [Mg/Fe]-[Fe/H] relations from both the L025m6 and L025m7 models agree and are insensitive to the assumed diffusion coefficient. For [Fe/H]${<}-1$, the high diffusion model shows good numerical convergence, while models with smaller diffusion coefficients converge less well. This suggests that when we increase the resolution of the simulations, metal mixing is enhanced in low-metallicity gas, leading to fewer iron-poor particles with low [Mg/Fe] values. Therefore, to achieve weak convergence, the metal diffusion coefficient may have to decrease with increasing resolution.

Observations suggest that the mean [Mg/Fe] for stars with $\rm{[Fe/H]}<-1$ ranges between 0.2 and 0.4. The L025m6 simulation with default diffusion coefficient ($C_{\rm{diffusion}}=0.01$) appears to match the data best. We adopt this model as our default for two reasons: it produces moderate metal mixing and therefore contains many metal-poor star particles with [Mg/Fe]$<0.5$; and it matches the observed trend for $\rm{[Fe/H]}<-2$. However, we caution that elemental abundances of low-metallicity gas may be sensitive to the poorly constrained turbulent diffusion coefficient, particularly in lower resolution simulations, where stochastic sampling of enrichment events also play an important role.

\section{Results}\label{Sec5}

\subsection{Stellar mass -- metallicity relation}

\begin{figure*} 
\begin{center}
\includegraphics[angle=0,width=0.45\textwidth]{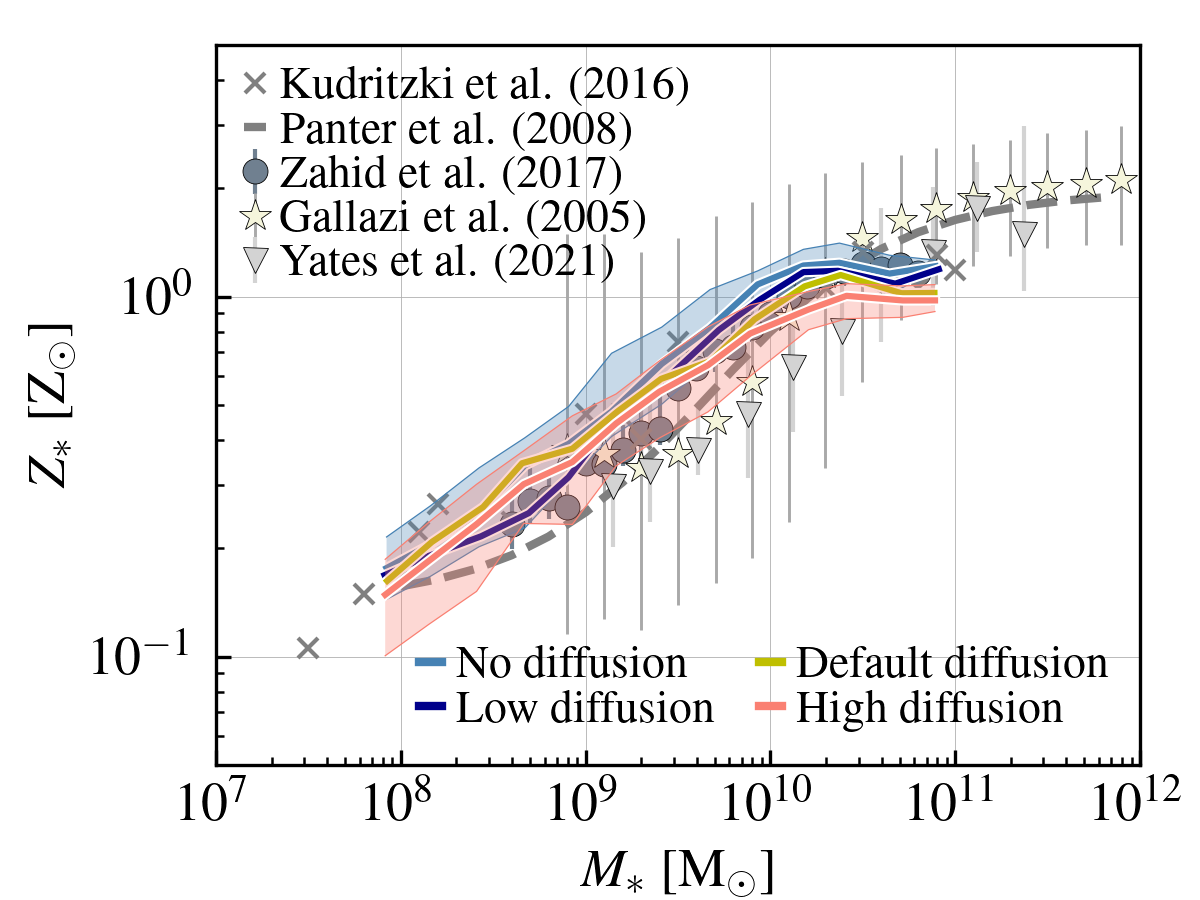}
\includegraphics[angle=0,width=0.45\textwidth]{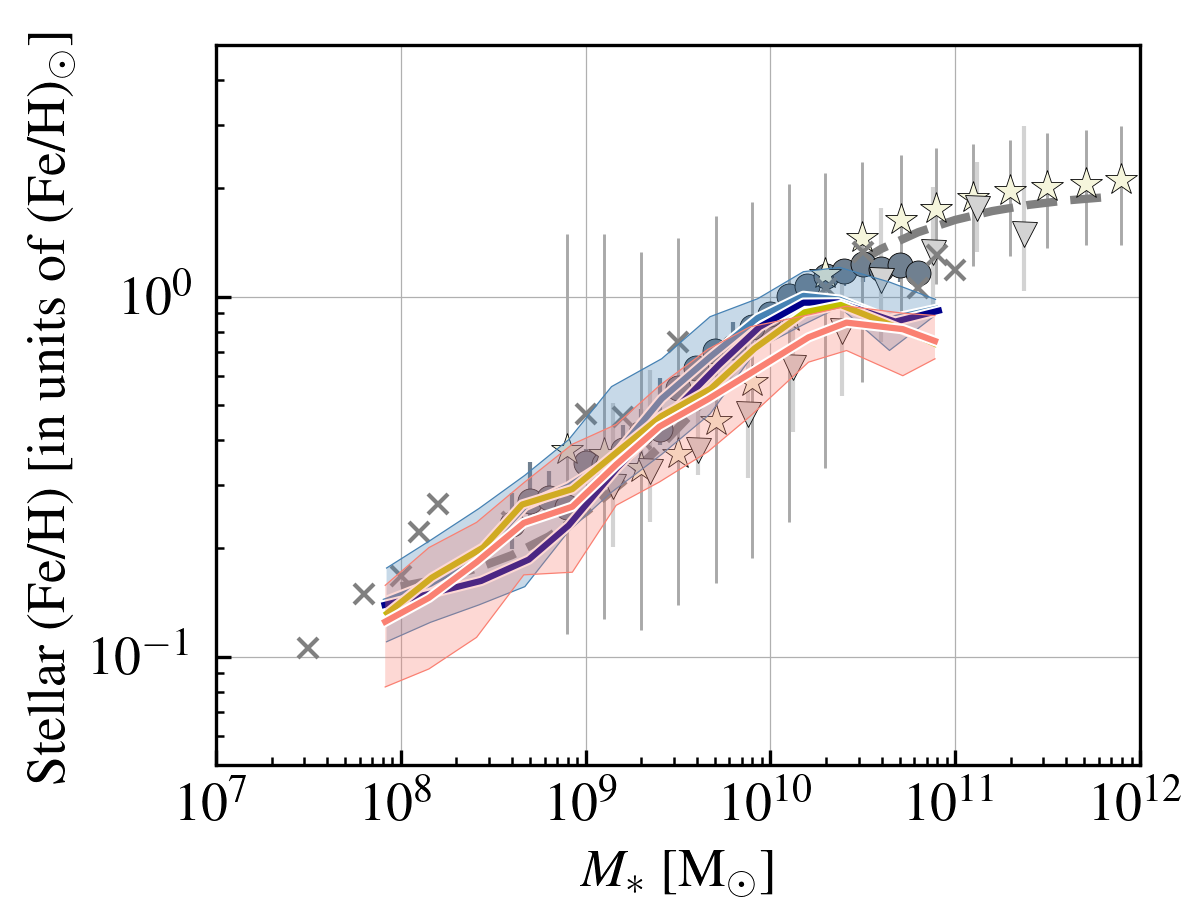}
\caption{Total stellar metallicity (left panel) and total stellar iron-to-hydrogen ratios (right panel) as a function of stellar mass. The data in the panels is shown relative to solar assuming $Z_{\odot}=0.0134$ and $\log_{10}(n_{\rm{Fe}}/n_{\rm{H}})_{\odot}+12=7.5$ (\citealt{Asplund09}). Curves show the median mass-weighted mean metallicities for the L025m6 simulations at $z=0$ with different diffusion coefficients, and the shaded regions show the 16th-84th percentiles. The panels show observations reported by \citet{Yates21}, \citet{Kudritzki16}, \citet{Zahid17}, \citet{Panter08}, and \citet{Gallazzi05} (all renormalized to solar abundances values from Asplund et al. 2009 for consistency).}
\label{MassMetallicityRelation}
\end{center}
\end{figure*}

A first step in the analysis of the simulations is to investigate the mass-metallicity relation. Fig.~\ref{MassMetallicityRelation} shows the mass-metallicity relation based on the total stellar metallicity inside central galaxies (left panel) and on the stellar iron-to-hydrogen ratios (right panel). The panels include relations from the L025m6 simulation under different diffusion coefficients: high ($C_{\rm{diffusion}}=0.1$, orange), default ($C_{\rm{diffusion}}=0.01$, light green), low ($C_{\rm{diffusion}}=0.001$, dark blue), and no diffusion ($C_{\rm{diffusion}}=0$, light blue). The curves mark the median mass-metallicity relation from the simulations at $z=0$ based on the mass-weighted mean stellar metallicity within a 3D 50 kpc aperture (left panel), and the mass-weighted iron-to-hydrogen ratio (right panel). The 0.2 dex stellar mass bins shown are those that include at least 10 galaxies per bin. The curves include galaxies that contain more than 100 stellar particles, which corresponds to a minimum stellar mass of $M_{*}\approx 10^{8}~\rm{M}_{\odot}$. The median curves include all central galaxies from the simulations, irrespective of whether they are star-forming or not. For a detailed comparison of the mass–metallicity relation for star-forming galaxies, the reader is referred to Figures 20 and 21 in \citet{Schaye25}.

The left panel shows that the stellar mass-metallicity relation is mildly impacted by diffusion. For galaxies with $M_{*}< 10^{9.5}~\rm{M}_{\odot}$, the impact of diffusion on the total metallicity is negligible compared with the scatter, and the small differences are not monotonic with the value of the diffusion coefficient. For higher mass galaxies, the metallicity decreases systematically with the value of the diffusion coefficient. For example, the median metallicity at $10^{10}~\rm{M}_{\odot}$ in the no difffusion model is $1.11~\rm{Z}_{\odot}$, whereas in the high-diffusion model it decreases to $0.82~\rm{Z}_{\odot}$. The scatter in the relation is comparable across all models, ranging between 0.1 and 0.2 dex. 

The left panel also shows the galaxy stellar metallicity increases with stellar mass until galaxies reach masses of $10^{10}$ M$_{\odot}$. Above this mass all models predict a plateau with galaxies having nearly constant solar metallicity regardless of their mass. This turnover mass and flattening agrees with the observational data from \citet{Zahid17} and \citet{Kudritzki16}, who fitted the stacked spectra from star-forming galaxies in SDSS with stellar population synthesis models to derive metallicities. Other datasets, such as \citet{Yates21}, who used a galaxy sample from the MaNGA FIREFLY value-added catalogue (\citealt{Wilkinson17,Goddard17}), which provided absorption-line based stellar metallicities obtained using the FIREFLY spectral-fitting code, do not show a flattening above the $10^{10}$ M$_{\odot}$ mass-scale. Instead, the \citet{Yates21} metallicities flatten off for galaxies more massive than $10^{11}$ M$_{\odot}$.

The right panel of Fig.~\ref{MassMetallicityRelation} shows the mass-weighted iron-to-hydrogen ratio in units of solar. This metallicity indicator follows the same trend as the total metallicity shown in the left panel. The stellar [Fe/H]-$M_{*}$ relation is mildly impacted by diffusion, with massive galaxies having a larger [Fe/H] ratio in models with lower diffusion coefficient.

\subsection{Stellar mass -- gas phase oxygen relation}

\begin{figure*} 
\begin{center}
\includegraphics[angle=0,width=0.45\textwidth]{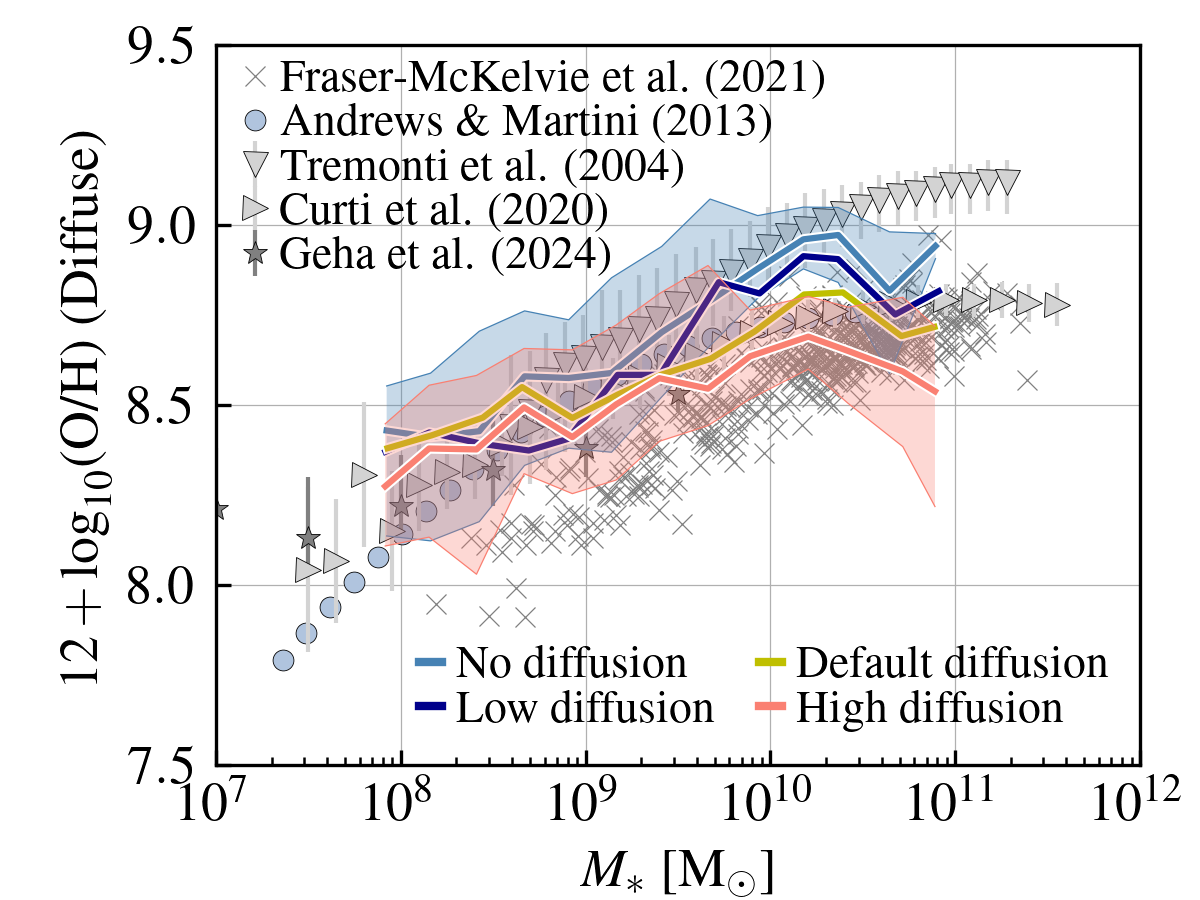}
\includegraphics[angle=0,width=0.45\textwidth]{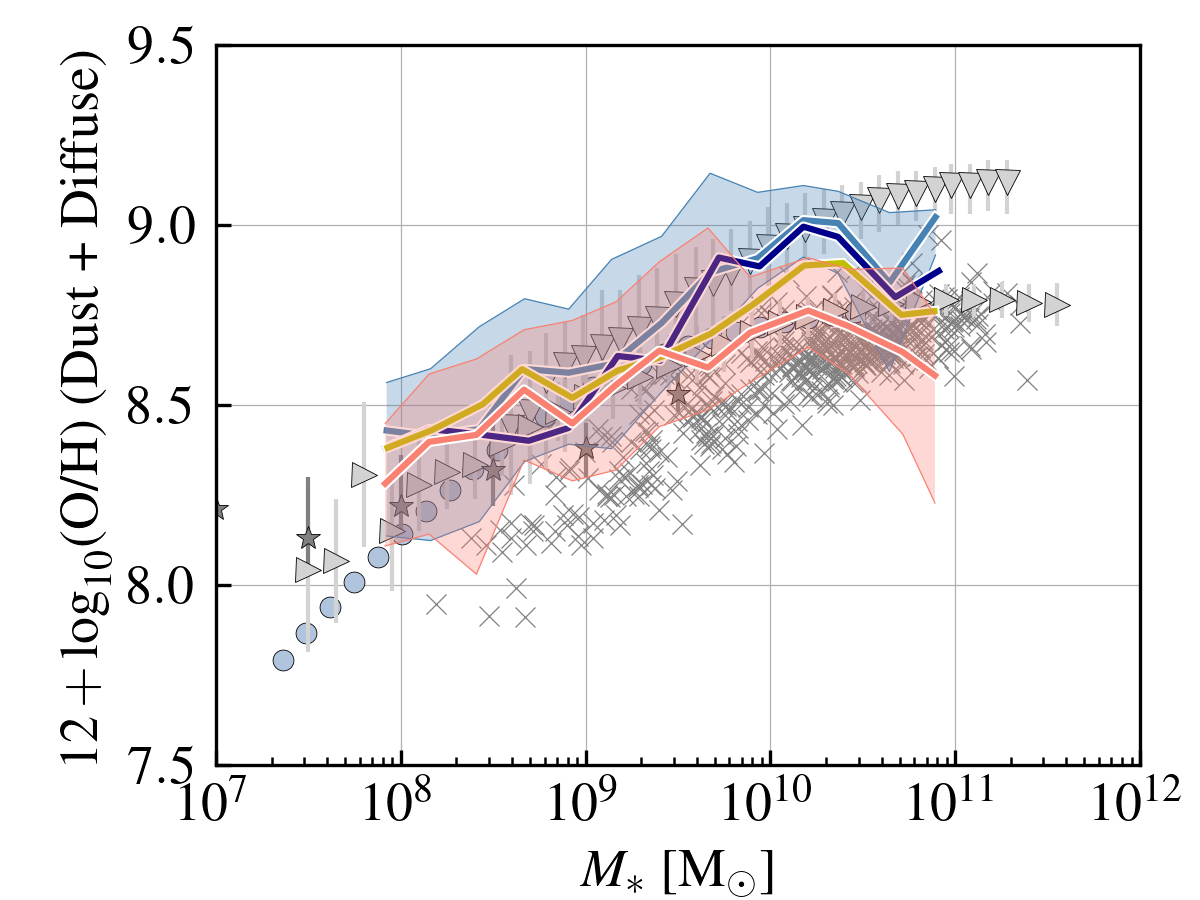}
\caption{Gas metallicity of the ISM as a function of stellar mass for undepleted (excluding metals present in dust, left panel) and including metals present in dust (right panel). Similar to Fig.~\ref{MassMetallicityRelation}, curves show the median mass-weighted mean metallicity for the L025m6 simulations at $z=0$ including all galaxies (regardless of whether they are star forming). The shaded regions show the 16th-84th percentiles. We only use cool, dense gas ($T<10^{4.5}$ K, $n_{\rm{H}}>0.1$ cm$^{-3}$, without imposing a minimum in metallicity) in the estimation of the galaxies' gas-phase abundance. The data points show observations reported by \citet{Geha24}, \citet{Fraser22}, \citet{Curti20}, \citet{Andrew13} and \citet{Tremonti04}.}
\label{MassGasMetallicityRelation}
\end{center}
\end{figure*}

We analyze the total metallicity of the ISM as a function of galaxy stellar mass. The ISM is defined by all gravitationally-bound, cool, dense gas particles ($T < 10^{4.5}$ K and $n_{\rm{H}} > 0.1$ cm$^{-3}$) within a 50 kpc aperture. Fig.~\ref{MassGasMetallicityRelation} presents the mass-weighted mean oxygen-to-hydrogen ratios, O/H, calculated as

\begin{equation}\label{OHeq}
{\rm{O/H}}=\frac{m_{\rm{H}}}{m_{\rm{O}}}\frac{\sum_{i}(X_{\rm{O}}/X_{\rm{H}})_{i}m_{{\rm{gas}},i}}{\sum_{i}m_{{\rm{gas}},i}},
\end{equation}

\noindent with $m_{\rm{H}}/m_{\rm{O}}$ the ratio of masses of hydrogen and oxygen nuclei, $X_{k}$ the element mass fraction, and $m_{{\rm{gas}},i}$ the mass of gas particle $i$. These ratios are calculated for both undepleted gas metallicity (left panel, labeled "diffuse", which excludes metals in dust) and total gas metallicity (right panel, labeled "diffuse + dust", which includes metals in dust). As in Fig.~\ref{MassMetallicityRelation}, each curve corresponds to the L025m6 simulations using varying diffusion coefficients.

The panels in Fig.~\ref{MassGasMetallicityRelation} also show observational datasets from \citet{Tremonti04}, \citet{Andrew13}, \citet{Curti20}, \citet{Fraser22}, and \citet{Geha24}. Tremonti et al. derived oxygen abundances for star-forming SDSS galaxies using statistical fits of multiple emission lines, incorporating a photoionization model which includes the depletion of heavy elements onto dust grains and the absorption of ionizing photons by dust. \citet{Andrew13} used a direct method based on the electron temperature measured from stacked SDSS galaxies, enhancing emission line signals by grouping galaxies by stellar mass and SFR. \citet{Curti20} also studied SDSS galaxies' oxygen abundances using a strong-line diagnostic combination. \citet{Fraser22} analyzed galaxies from the SAMI survey with a direct method. \citet{Geha24} provided median gas-phase metallicities from satellite galaxies from the SAGA survey using the strong line $\rm{N}_{2}$ diagnostic (\citealt{Kewley08}). Notably, the mass-metallicity relations from these studies vary significantly, particularly between \citet{Tremonti04} and \citet{Fraser22}, likely due to differences in the calibration of the emission-line diagnostics (\citealt{Kewley08}, and/or underestimations from a non-negligible fraction of metals condensed onto dust-grains (\citealt{Mattsson12}).

Fig.~\ref{MassGasMetallicityRelation} shows that diffusion has a greater impact on $z=0$ gas-phase oxygen abundance than on $z=0$ stellar metallicity. For galaxies with stellar masses larger than $10^9~\rm{M}_{\odot}$ the metallicity decreases systematically with the value of the diffusion coefficient, with differences increasing towards higher stellar masses. For instance, $10^{10}~\rm{M}_{\odot}$ galaxies have $12+\log_{10}(\rm{O}/\rm{H}) = 8.9$ in the no-diffusion model, versus 8.6 in the high-diffusion model. Weaker diffusion results in a steeper mass-metallicity relation. The default diffusion model, chosen based on its agreement with [Mg/Fe] of metal-poor stars, aligns well with the relations from \citet{Curti20} and \citet{Andrew13}.

Since the fraction of metals depleted onto dust grains is not well-constrained observationally, we show both undepleted and total gas metallicities in Fig.~\ref{MassGasMetallicityRelation} for a fair comparison. The right panel shows that accounting for metals in dust raises the mass-weighted oxygen-to-hydrogen ratio, while the trend with the diffusion coefficient remains unchanged. For further discussion of the influence of depletion from the perspective of the COLIBRE dust model, see section 3.8 of \citet{Trayford25}. 

The trend observed with metal diffusion in Fig.~\ref{MassGasMetallicityRelation} can be attributed to the smoothing of metallicity gradients between the inner and outer regions of the galaxy, and/or between the ISM and the circumgalactic medium. This overall mixing leads to a reduction in the median metallicity of the galaxy.

\subsection{Carbon, nitrogen and oxygen}

In this section, we analyze the relation between gas-phase metallicity $12+\log_{10}(\rm{O}/\rm{H})$, nitrogen-to-oxygen ratio [N/O], and carbon-to-oxygen ratio [C/O]. These ratios provide insights into the physical processes and timescales involved in the production of the different elements. Carbon is synthesized during helium burning via the triple-$\alpha$ process in main sequence and giant branch stars. Oxygen is then produced by the fusion of carbon with additional $\alpha$ particles and released into the ISM through core-collapse supernovae (CCSNe). Nitrogen has both primary and secondary production channels. Primary nitrogen mainly originates in low- and intermediate-mass stars, though it can also form in massive stars with rotational mixing (e.g., \citealt{Vincenzo16}). Secondary nitrogen is a byproduct of the CNO cycle, depending on the pre-existing amounts of C and O within the star.

The observed [N/O] vs. [O/H] relation follows a well-established trend: at low metallicities, [N/O] remains roughly constant around $\rm{[N/O]}\approx -1.5$. As metallicity increases, [N/O] rises rapidly with [O/H]. This low-metallicity plateau in [N/O] is interpreted as evidence for a primary nitrogen component from massive stars (\citealt{Chiappini03,Chiappini05,Vincenzo16}). Similarly, [C/O] shows a constant value of about [C/O]$\approx -0.8$ at low metallicities, with [C/O] increasing alongside [O/H] at higher metallicities (\citealt{Nicholls17}).

\begin{figure*} 
\begin{center}
\includegraphics[angle=0,width=0.8\textwidth]{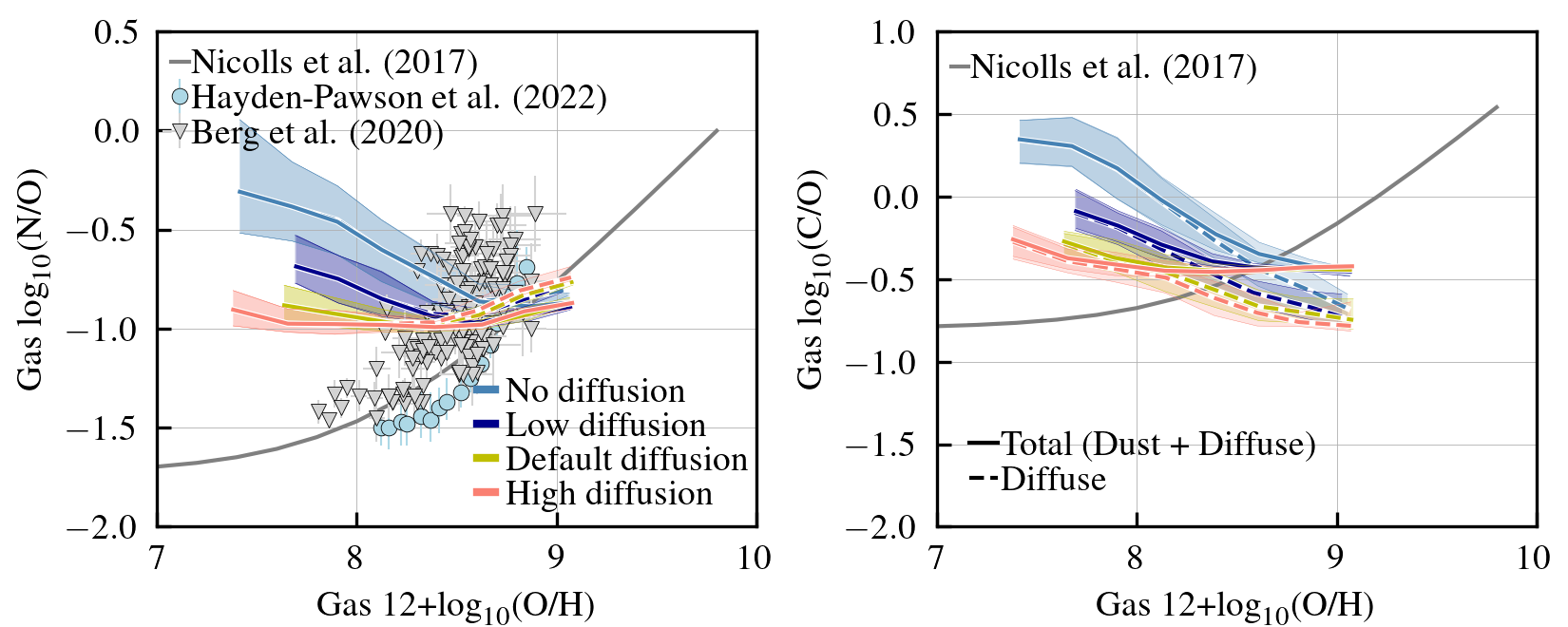}
\caption{Gas-phase nitrogen-to-oxygen ratio (left panel) and carbon-to-oxygen ratio (right panel) as a function of the gas-phase metallicity $12+\log_{10}(\rm{O}/\rm{H})$. The different coloured lines depict the median trends at $z=0$ from L025m6 simulations with different diffusion coefficients. In the calculation of the galaxies' mass-weighted mean gas-phase ratios ([O/H], [N/O] and [C/O]), we consider both the undepleted gas (shown by dashed lines and labelled as diffuse) and the total gas plus dust (shown by solid lines and labelled diffuse + dust). The panels include gas-phase abundance data from \citet{Nicholls17} (based on calibrations from Milky Way stars), \citet{Hayden22} (based on gas metallicity estimations from SDSS galaxies) and \citet{Berg20} (based on gas-phase abundances from H~\textsc{ii} regions from four spiral galaxies).}
\label{GasAbundance_cno}
\end{center}
\end{figure*}

Fig.~\ref{GasAbundance_cno} illustrates the [N/O] vs. [O/H] relation (left panel) and the [C/O] vs. [O/H] relation (right panel). Both panels include gas-phase scaling relations from \citet{Nicholls17} (grey solid lines), based on Milky Way stellar abundances converted to oxygen-based nebular metallicity. The blue circles indicate the median trend from \citet{Hayden22} for gas-phase abundances from SDSS galaxies, while the data points from \citet{Berg20} depict direct-method gas-phase abundances for H~\textsc{ii} regions across four spiral galaxies.

Fig.~\ref{GasAbundance_cno} also shows the median mass-weighted mean abundance relations from star-forming galaxies (i.e. with non-zero SFR) with masses above $10^{8}~\rm{M}_{\odot}$ in the L025m6 simulations. As in Section 5.2, we define the ISM as cool, dense gas particles ($T<10^{4.5}$ K, $n_{\rm{H}}>0.1$ cm$^{-3}$) within a 50 kpc aperture. We calculate the mass-weighted mean gas-phase ratios ([O/H], [N/O], and [C/O], as in eq.~(\ref{OHeq})) for both undepleted gas (shown with dashed lines, labeled ``diffuse'') and total gas plus dust (solid lines, labeled ``diffuse + dust''). The panels display median relations from the simulations using different diffusion coefficients: high ($C_{\rm{diffusion}}=0.1$, orange), default ($C_{\rm{diffusion}}=0.01$, light green), low ($C_{\rm{diffusion}}=0.001$, dark blue), and no diffusion ($C_{\rm{diffusion}}=0$, light blue).

Fig.~\ref{GasAbundance_cno} highlights the impact of diffusion on the median abundance relation. Lower diffusion rates correspond to higher [N/O] and [C/O] ratios at low metallicities ($12+\log_{10}(\rm{O}/\rm{H})<8.5$). At higher metallicities, the median trends converge, with the diffuse gas showing a slight increase in [N/O] as [O/H] increases (left panel), while [C/O] decreases with increasing [O/H] (right panel). However, it can be seen that regardless of the diffusion model, the relative abundance relations from the simulations do not match the observations. We also tested the median relations of the individual star-forming gas in massive galaxies, rather than the mass-weighted mean trend, but found that the relations remained unchanged. In the default diffusion model, the median total [N/O] and [C/O] ratios stay relatively constant as [O/H] increases.

One possible way to improve agreement with the observed [N/O]-[O/H] relation is by adjusting the yields. We found that adding a CCSN nitrogen boost factor of 1.5 and reducing low-metallicity ($Z<10^{-2}$) AGB yields of nitrogen by a factor of 0.5 to 0.2, improves the agreement in the following relations: stellar [N/Fe] vs. [Fe/H], gas-phase [N/O] vs. [O/H], and gas-phase [N/O] vs. stellar mass (not shown), matching observations from APOGEE and \citet{Hayden22}. However, these changes appear quite ad hoc, so we chose not to implement them. Further exploration of these relations motivates a future analysis of the metallicity-dependence of the adopted yields. Appendices \ref{AGBNucleosynthesis_Sec} and \ref{CCSNNucleosynthesis_Sec} present further details on the mass and metallicity dependence of the AGB/CCSN yields used in this study.

\subsection{Barium, strontium and europium}

This section resumes the analysis presented in Section \ref{CCSN_boost_factors_section}, where we analysed the stellar abundance ratios from 27 Milky Way-like galaxies from the L025m6/Default diffusion simulation based on their halo mass, stellar mass, and disc morphology. From these galaxies, we select all gravitationally-bound star particles (without imposing a distance cut) in order to compare their abundance ratios with those of stars in the Milky Way’s disc and halo.

\begin{figure*} 
\begin{center}
\includegraphics[angle=0,width=\textwidth]{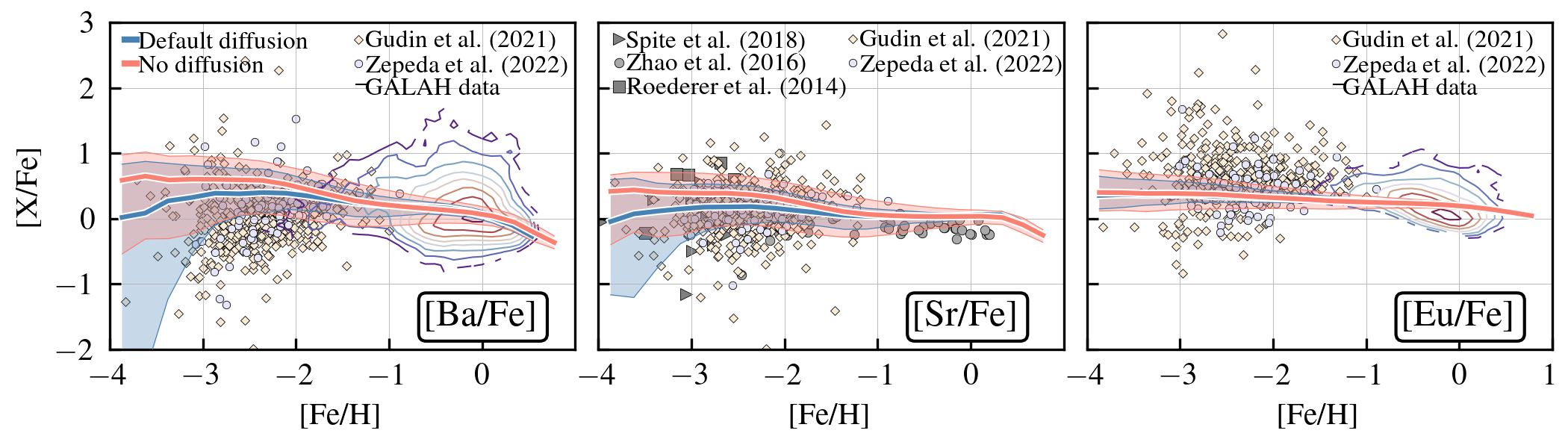}
\caption{Element abundance ratios from simulated star particles are compared to observed abundances of the Milky Way stars. Each panel shows the abundance relative to iron of a different element: barium (left panel), strontium (middle panel), and europium (right panel), all plotted as a function of iron over hydrogen. Solid coloured lines show the $z=0$ median relations from L025m6 simulations with different diffusion coefficients, while shaded regions mark the 16-84th percentiles. The contours show Milky Way stellar abundances from the GALAH survey (\citealt{Buder21}). Additional observational data corresponds to \citet{Zepeda22} (grey circles), \citet{Gudin21} (light diamonds), \citet{Spite18} (grey triangles), \citet{Zhao16} (dark grey circles), and \citet{Roederer14} (grey squares).}
\label{StellarAbundance_rs_processes}
\end{center}
\end{figure*}

Fig.~\ref{StellarAbundance_rs_processes} shows the abundance relative to iron of barium (left panel), strontium (middle panel), and europium (right panel) as a function of iron over hydrogen. Solid lines represent the median relations of the stellar particles of all galaxies combined, while shaded regions mark the 16-84th percentiles. The panels display median relations from the simulations using different diffusion coefficients: default ($C_{\rm{diffusion}}=0.01$, blue) and no diffusion ($C_{\rm{diffusion}}=0$, red). For clarity, the high and low diffusion models are omitted, as their results overlap with the default model.

The panels also include a compilation of observational datasets. Coloured contours correspond to the Milky Way stellar abundances from the GALAH survey (\citealt{Buder21}). Light diamonds correspond to data from \citet{Gudin21} for metal-poor stars in the halo and disk of the Milky Way. Light grey circles correspond to metal-poor halo stars from \citet{Zepeda22}. Grey triangles, circles and squares correspond to \citet{Spite18}, \citet{Zhao16}, and \citet{Roederer14}, respectively. 

It can be seen from the panels of Fig.~\ref{StellarAbundance_rs_processes} that diffusion does not have a large impact on the median trend of the abundance ratios as a function of iron. For barium and strontium, elements that are produced exclusively by AGB stars, diffusion increases the scatter in the relation, with very metal poor star particles having lower abundance ratios in the default diffusion model relative to the no diffusion model. Europium as a function of iron remains roughly unchanged when diffusion is switched off. This indicates that all star particles are equally enriched with europium by neutron star mergers. We find that neutron stars are the primary contributors of europium, whereas CEJSN and collapsars, due to their low europium yields and small event rate, produce a very small contribution. 


Fig.~\ref{StellarAbundance_rs_processes} shows that our default model successfully reproduces the observed median trends of both $r$-process and $s$-process elements as a function of iron. However, the model does not reproduce the observed scatter in these relations, nor is it expected to, since the model predicts abundances of stellar populations rather than those of individual stars.

\citet{vandeVoort22} explore a scenario that may allow our model to better reproduce the scatter in the europium-iron relation: the inclusion of natal kicks. Some neutron stars seem to have velocities exceeding 1000 km s$^{-1}$ (\citealt{Hobbs05}). This is not observed in their progenitor population, which means that their high velocities may be imparted during the supernova explosions, that expel the outer material in an asymmetric manner, causing the neutron stars to be born with a significant kick velocity. \citet{vandeVoort22} demonstrated that the inclusion of additional massless neutron star merger particles which are given a velocity kick in a random direction, increases the scatter of europium in metal-poor star particles. While this result motivates future exploration of additional europium sources and their spatial distribution in the ISM, it remains premature to conclude that features such as natal kicks are necessary, especially given the current resolution limits that prevent modeling individual stars.

\begin{figure*} 
\begin{center}
\includegraphics[angle=0,width=\textwidth]{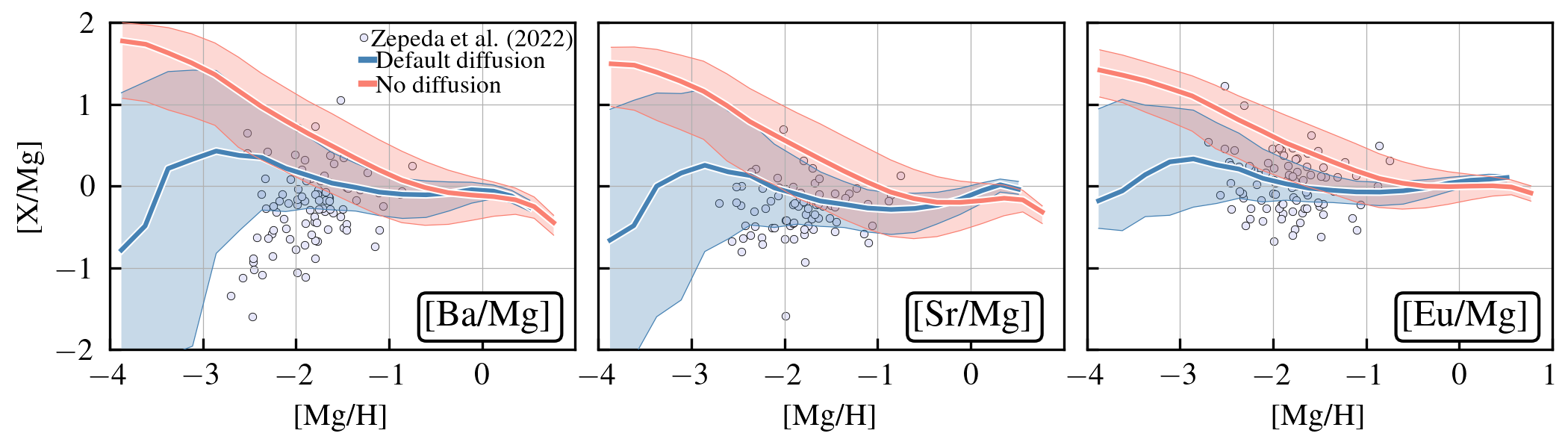}
\caption{Similar to Fig.~\ref{StellarAbundance_rs_processes}, but in this case each panel shows the abundance relative to magnesium of a different element: barium (left panel), strontium (middle panel), and europium (right panel), all plotted as a function of magnesium over hydrogen. The solid coloured lines show the median $z=0$ relations from L025m6 simulations with different diffusion coefficients, while shaded regions mark the 16-84th percentiles. Observational data corresponds to \citet{Zepeda22} (grey circles).}
\label{StellarAbundance_rs_Mg_processes}
\end{center}
\end{figure*}

Fig.~\ref{StellarAbundance_rs_Mg_processes} also shows the abundance ratios relative to magnesium of barium (left panel), strontium (middle panel), and europium (right panel), but in this case all ratios are calculated as a function of magnesium over hydrogen. It is interesting to see that for these relations, diffusion has a large impact. The abundance ratios decrease with increasing diffusion at fixed magnesium over hydrogen provided [Mg/H]$\lesssim-1$. This is essentially driven by magnesium. Diffusion increases magnesium abundances for most low-metallicity particles, as CCSN products are relatively poorly sampled and magnesium is produced almost exclusively ($\approx$90\%) by CCSNe (Fig.~\ref{StellarAbundance}).

\subsection{General relations}

\begin{figure*} 
\begin{center}
\includegraphics[angle=0,width=0.7\textwidth]{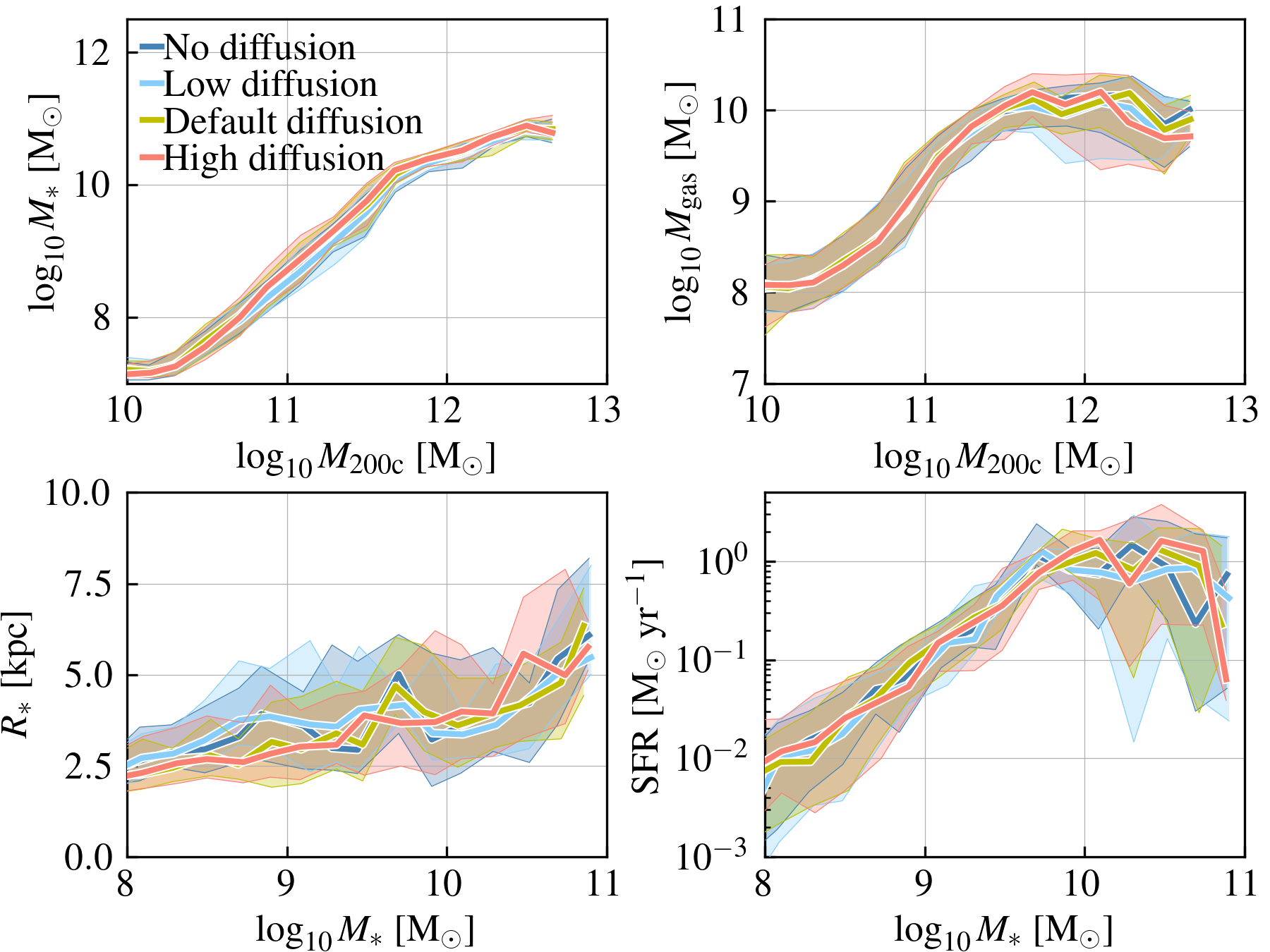}
\caption{General galaxy scaling relations at $z=0$: median stellar mass-halo mass (top left), gas mass-halo mass (top right), 3D half mass radius-stellar mass (bottom left), and SFR-stellar mass (bottom right). We have considered all central galaxies more massive than $10^{8}$ M$_{\odot}$ in stellar mass. Solid lines indicate median relations with shaded regions marking the 16th and 84th percentiles. Metal diffusion has only a minimal impact on these basic galaxy properties.}
\label{Generalrelations}
\end{center}
\end{figure*}

In this final section we investigate whether metal diffusion has an impact on the basic properties of central galaxies. For this we inspect some general scaling relations, namely the stellar mass-halo mass relation, gas mass-halo mass relation, 3D stellar half mass radius-stellar mass relation, and SFR-stellar mass relation for central galaxies at $z=0$. These relations are shown in Fig.~\ref{Generalrelations}, where each panel depicts the median trends from the L025m6 simulations using different diffusion coefficients: high ($C_{\rm{diffusion}}{=}0.1$, orange), default ($C_{\rm{diffusion}}{=}0.01$, light green), low ($C_{\rm{diffusion}}{=}0.001$, light blue), and no diffusion ($C_{\rm{diffusion}}{=}0$, dark blue). In the top panels, the stellar mass and gas mass are defined by summing up all star and gas particles within a 3D aperture with radius of 50 kpc. The halo mass is defined as the total mass of all particles enclosed within the radius at which the mean internal density is 200 times the critical density of the Universe. The bottom right panel of the figure shows the SFR calculated by summing the SFR of all gas particles within a 50 kpc aperture. For this panel, we only included star-forming galaxies, defined throughout this work as those galaxies with non-zero SFR. Fig.~\ref{Generalrelations} does not show observational data, we refer the reader to \citet{Schaye25} and \citet{Chaikin25} for comparisons with observations. Here, we only intend to illustrate the impact of metal diffusion on these galaxy properties.

Fig.~\ref{Generalrelations} shows that metal diffusion has a minimal impact on the general galaxy properties. While the top left panel indicates a small increase ($\sim 0.1$ dex) of the median stellar mass in the high diffusion model relative to the low diffusion model in $~10^{11}~\rm{M}_{\odot}$ haloes, this difference lies within the scatter. The bottom left panel also shows a slight increase in the galaxies' radii at masses of $~10^{9}~\rm{M}_{\odot}$ in the no diffusion and low diffusion models relative to the rest. However, this difference is within the scatter of the relation. 

\section{Conclusions}\label{Sec6}

We have presented the chemical enrichment module of the COLIBRE subgrid model of galaxy formation. By creating an updated set of nucleosynthetic yield tables and modelling the evolution of 12 chemical elements through multiple enrichment channels--- including CCSNe, SNIa, and AGB stars--- we have developed a comprehensive framework to investigate the chemical evolution of galaxies.

The model explicitly includes the production of $s$-process elements (barium and strontium) from AGB stars and $r$-process elements (europium) from neutron star mergers, common envelope jet supernovae, and collapsars. Furthermore, the enrichment module incorporates metal mixing as a turbulent diffusion process. This is motivated by the Smagorinsky model of turbulent cascade, where the diffusion coefficient is tied to the local velocity shear, providing a better treatment of mixing in shearing flows.

There are free parameters in our model. These include CCSN boost factors, which compensate for yield uncertainties that persist even under fixed IMFs and when neglecting stellar rotation and binary effects. We adopt a fiducial boost factor of 1 for most elements, i.e. no boost factor, but we increase the carbon and magnesium CCSN yields by a factor of 1.5 in order to match key stellar abundance ratios, specifically [X/Fe] as a function of [Fe/H], from the APOGEE DR14 survey (Fig.~\ref{StellarAbundance}). The model also successfully reproduces observed $\alpha$-enhancement trends with metallicity from SDSS galaxies, and with stellar mass for early-type galaxies from the ATLAS-3D survey and dwarf galaxies from the Local Group (Fig.~\ref{alphaenhancement}).

Another tunable parameter is the diffusion coefficient, set to $C_{\rm diffusion} = 0.01$ to reproduce the observed [Mg/Fe]–[Fe/H] relation for metal-poor stars ($\rm{[Fe/H]} < -1$), as indicated by data from \citet{Venn04}, \citet{Yong13}, and the Pristine Survey (Fig.~\ref{DiffusionAnalysisPlot}).

We find that diffusion has negligible effects on fundamental $z=0$ galaxy scaling relations such as the stellar mass–halo mass, gas mass–halo mass, size–mass, and SFR–mass relations (Fig.~\ref{Generalrelations}). Its impact on the stellar mass–metallicity relation is mild, with slightly lower [Fe/H] values in massive galaxies when diffusion is included (Fig.~\ref{MassMetallicityRelation}). However, the gas-phase metallicity is more strongly affected: for galaxies with $M_{*} > 10^9~\mathrm{M}_\odot$, increased diffusion leads to lower oxygen abundances, with the offset becoming more pronounced in higher-mass systems (Fig.~\ref{MassGasMetallicityRelation}). This behaviour likely reflects the smoothing of metallicity gradients within galaxies and/or between the interstellar and circumgalactic media.

We have also examined the gas-phase C/O and N/O ratios. Across all diffusion models tested, the simulations fail to reproduce the observed trends in these abundance ratios (Fig.~\ref{GasAbundance_cno}), highlighting the need for improved treatments of nitrogen and carbon yields.

Lastly, we analyse the stellar abundances of $s$- and $r$-process elements, and show that our model reproduces the median observed trends as a function of both iron and magnesium (Figs.~\ref{StellarAbundance_rs_processes} and \ref{StellarAbundance_rs_Mg_processes}).

Looking ahead, there are several directions for future improvements. One key area is the refinement of yield tables, including the effects of stellar rotation, particularly for nitrogen and carbon (Fig.~\ref{GasAbundance_cno}). Also for the less well-understood processes like neutron star mergers and collapsars, the accuracy of heavy-element predictions could be improved in concert with new and upcoming observational insights. In addition, coupling the chemical enrichment approach presented here with a model for variable stellar initial mass functions (IMFs) could provide powerful insight into how IMF variations influence galactic chemical evolution.

\section*{Acknowledgements}

JT acknowledges support of a STFC Early Stage Research and Development grant (ST/X004651/1). SP acknowledges funding by the Austrian Science Fund (FWF), grant number V 982-N. ABL acknowledges support by the Italian Ministry for Universities (MUR) program “Dipartimenti di Eccellenza 2023-2027” within the Centro Bicocca di Cosmologia Quantitativa (BiCoQ), and support by UNIMIB’s Fondo Di Ateneo Quota Competitiva (project 2024-ATEQC-0050). This project has received funding from the Netherlands Organization for Scientific Research (NWO) through research programme Athena 184.034.002. CSF acknowledges support from the European Research Council (ERC) Advanced
Investigator grant DMIDAS (GA 786910) and the STFC Consolidated Grant ST/T000244/1. This work used the DiRAC Durham facility managed by the Institute for Computational Cosmology on behalf of the STFC DiRAC HPC Facility (www.dirac.ac.uk). The equipment was funded
by BEIS capital funding via STFC capital grants ST/K00042X/1, ST/P002293/1, ST/R002371/1 and ST/S002502/1, Durham University and STFC operations grant ST/R000832/1. DiRAC is part of the National e-Infrastructure.

\section*{Data availability}
A public version of the SWIFT code (\citealt{Schaller23}) is available at \url{http://www.swiftsim.com}. The COLIBRE modules implemented in SWIFT will be made publicly available after the public release of the simulation data. The COLIBRE yield tables are available at \url{https://github.com/correac/COLIBRE_yields_tables.git}.

\bibliography{biblio}
\bibliographystyle{mnras}

\appendix

\section{AGB Nucleosynthesis yields}\label{AGBNucleosynthesis_Sec}

Section~\ref{AGB_yields_sec} describes the AGB nucleosynthesis yields that we adopt from \citet{Karakas10}, \citet{Fishlock14}, \citet{Karakas16}, \citet{Doherty14} and \citet{Cinquegrana22}, with table~\ref{Yield_tables} indicates the mass and metallicity ranges used from each work. In this section we explain how we integrate these datasets and create a look-up ``net stellar yield" table to be used by the COLIBRE code. For the interested reader, the full compilation of nucleosynthesis yields adopted by this work is publicly available in this github repository: $\rm{https://github.com/correac/COLIBRE\_yield\_tables}$.

Net yields, $Y_{k,\rm{tbl}}(M_{\rm{tbl}},Z_{\rm{tbl}})$ (in solar mass units, with the subscript ``tbl'' denoting tabulated values), are defined as the total mass of a star with initial mass $M_{\rm{tbl}}$ and metallicity $Z_{\rm{tbl}}$ that is converted into the element $k$ and returned to the ISM during its entire lifetime, $\tau_{\rm{tbl}} (M_{\rm{tbl}},Z_{\rm{tbl}})$. These stellar yields are calculated as follows

\begin{equation}
Y_{k,\rm{tbl}}(M_{\rm{tbl}},Z_{\rm{tbl}})=\int^{\tau_{\rm{tbl}}(M_{\rm{tbl}},Z_{\rm{tbl}})}_{0}[X_{k,\rm{tbl}}-X^{0}_{k,\rm{tbl}}]\dot{M}_{\rm{tbl}}{\rm{d}}t,
\end{equation}

\noindent where $\dot{M}_{\rm{tbl}}$ is the mass-loss rate of the star throughout its lifetime, and $X_{k,\rm{tbl}}$ and $X^{0}_{k,\rm{tbl}}$ refer to the current and initial abundance of the element $k$, respectively.

\citet{Karakas10}, \citet{Fishlock14}, \citet{Doherty14}, and \citet{Cinquegrana22} directly provide net yields. In the look-up tables we also include the total mass ejected, $M_{\rm{ej,total,tbl}}=\int^{\tau (M_{\rm{tbl}},Z_{\rm{tbl}})}_{0}\dot{M}_{\rm{tbl}}{\rm{d}}t=M_{\rm{final}}-M_{\rm{initial}}$, and the total amount of metals produced. \citet{Karakas16} only provide total yields, $Y_{k,\rm{tbl,total}}(M_{\rm{tbl}},Z_{\rm{tbl}})=\int^{\tau_{\rm{tbl}}(M_{\rm{tbl}},Z_{\rm{tbl}})}_{0}X_{k,\rm{tbl}}\dot{M}_{\rm{tbl}}{\rm{d}}t$, therefore for this dataset we calculate the net yields by substracting the initial abundance of element $k$ multiplied by the total mass ejected by the star: $Y_{k,\rm{tbl}}(M_{\rm{tbl}},Z_{\rm{tbl}})=Y_{k,\rm{tbl,total}}(M_{\rm{tbl}},Z_{\rm{tbl}})-X^{0}_{k,\rm{tbl}}\times M_{\rm{ej,total,tbl}}$.

It is possible that the net yields take negative values. In that case, they reflect the quantity of species $k$ that is destroyed inside the star over its lifetime (compared to the initial amount present in the wind). Conversely, positive yields quantify the production of species $k$ over the star's lifetime.

We expand the net yields from \citet{Karakas16} by incorporating data from \citet{Doherty14} through interpolation. Specifically, we extend the new yields for the mass range $8-9~\rm{M}_{\odot}$ and metallicity $Z=0.007$ and 0.014. For the $8-9~\rm{M}_{\odot}$ $Z=0.03$ metallicity bin, we use the $Z=0.02$ yields directly from \citet{Doherty14}. Linear extrapolation over the mass range $9-12$ M$_{\odot}$ is applied to finalize the new yields for the metallicity bins $Z=0.007$, 0.014 and 0.03.

For the metallicity bins $Z=10^{-4}$ and $Z=0.001$, we also linearly extrapolate the yields using the datasets from \citet{Karakas10} and \citet{Fishlock14} within the mass range $9-12$ M$_{\odot}$. In the case of the $Z=0.004$ net yields, we extend the \citet{Karakas10} data in the mass range $6-9~\rm{M}_{\odot}$ by incorporating the net yields from \citet{Doherty14} and linearly extrapolate towards the mass range $9-12$ M$_{\odot}$. Finally, for the metallicity range $Z=0.04-0.1$, net yields from \citet{Cinquegrana22} are linearly extrapolated within the mass range $8-12$ M$_{\odot}$.

\begin{figure*} 
\begin{center}
\includegraphics[angle=0,width=0.9\textwidth]{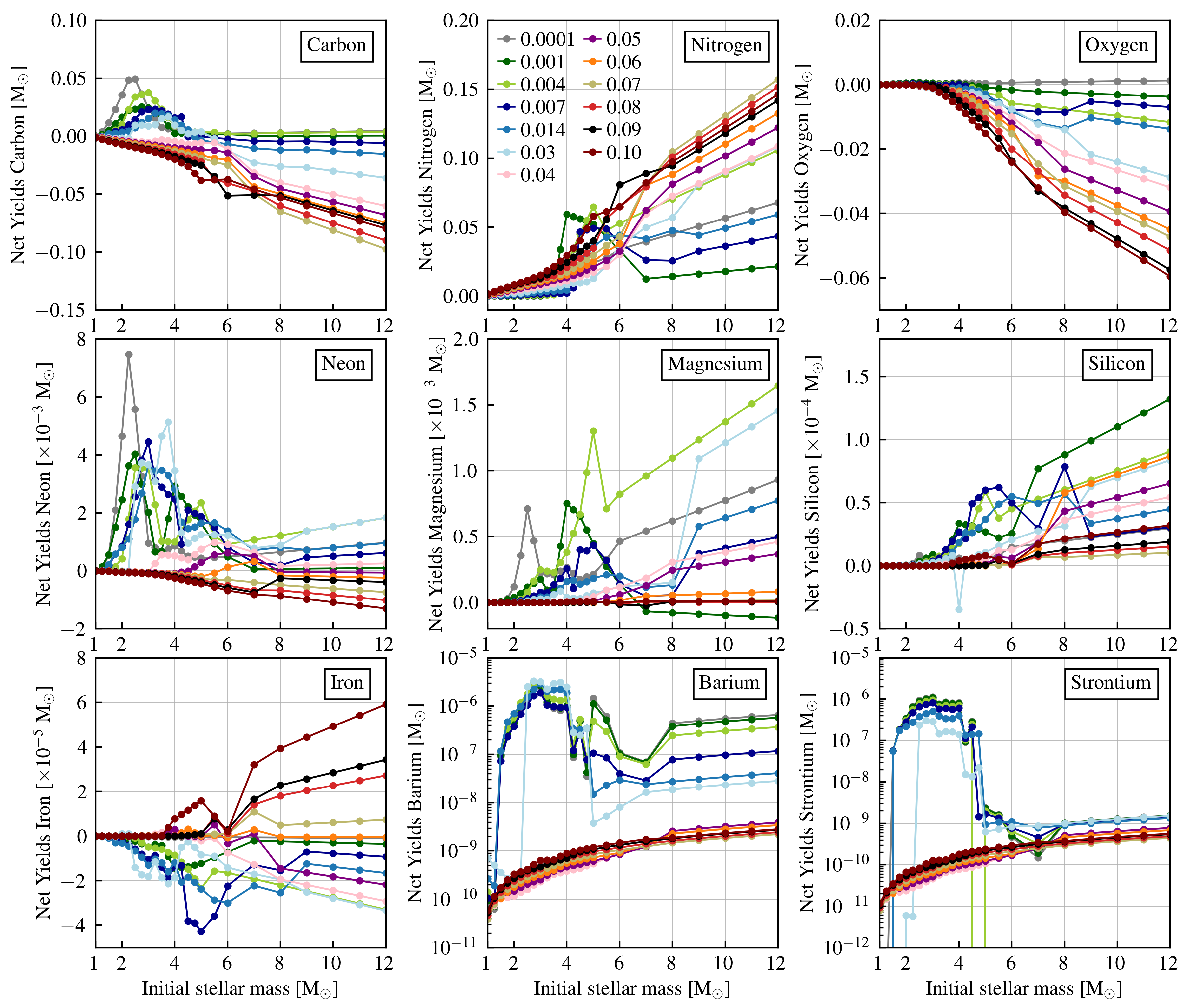}\\
\caption{Net yields from AGB stellar nucleosynthesis adopted in this work. Color lines correspond to different metallicity bins as indicated by the legend in the top middle panel.}
\label{AGByields}
\end{center}
\end{figure*}

Fig.~\ref{AGByields} shows the AGB net yields for all elements over the metallicity range $Z=0.0001-0.1$ and initial star mass of $1-12$ M$_{\odot}$.

\section{CCSN nucleosynthesis yields}\label{CCSNNucleosynthesis_Sec}

For CCSN nucleosynthesis yields we provide rather different look-up tables than those from AGB yields. We analyze the datasets from \citet{Kobayashi06} and \citet{Nomoto06}, presented as a function of the progenitor (zero-age main sequence) mass ($M_{\rm{tbl}}=$ 13, 15, 18, 20, 25, 30, and 40 $\rm{M}_{\odot}$) and metallicity ($Z_{\rm{tbl}}=$ 0, 0.001, 0.004, 0.02, and 0.05). From these we extract the total mass of element $k$ (or total metal mass) produced by the massive star and ejected during CCSNe event, $m_{{\rm{ej}},{\rm{total/metal}}/k}$. In addition, we calculate $m_{\rm{ej,tbl/winds}}$, defined as the total mass ejected by the massive star in the pre-supernova phase through winds. We calculate $m_{\rm{ej,tbl/winds}}=M_{\rm{initial}}-M_{\rm{final}}$, with $M_{\rm{initial}}$ the zero-age main sequence mass of the star and $M_{\rm{final}}$ the pre-supernovae phase mass.

Finally, we generate look-up tables containing $m_{{\rm{ej,tbl/CCSN}},k}(M_{\rm{tbl}}, Z_{\rm{tbl}})$, $m_{\rm{ej,tbl/winds}}(M_{\rm{tbl}}, Z_{\rm{tbl}})$, and the total mass ejected by the star, $m_{\rm{ej,tbl}}(M_{\rm{tbl}}, Z_{\rm{tbl}})$, during both the pre-supernova phase and the CCSN explosion, for stars within the mass range $6-40~\rm{M}{\odot}$.

\section{Metal conservation tests}

There is a potential issue with the methodology of metal diffusion described in Section~\ref{Metal_diffusion_Sec}. Eq.~\ref{4} indicates that the element abundances of a gas particle $i$ will be updated depending on the relative difference of the abundances of nearby gas particles. This in turn implies a direct subtraction and summation of small numbers, which, if properly handled, can produce the propagation of round off errors. The result can be the numerical loss of the gas particles total metal content. In addition to this effect, in SWIFT not all particles are ``active". SWIFT rewrites the classic leapfrog algorithm and considers sub-cycles where only a fraction of particles, those that are active, receive velocity updates whilst all other particles are only moved (drifted) to the current point in time of the simulations (see further details in \citealt{Schaller23}). As a result, particles are separated into active and inactive, depending on their time-step size. For the metal diffusion routine, only active particles exchange and receive metals from their neighboring particles and their metal content is updated accordingly alongside their velocity update. If a neighboring particle is inactive, it participates in the calculation of metal diffusion (via eq.~\ref{4}), but its metal content is not updated. In general, the code develops ``localized" time-steps regions, therefore neighbouring particles tend to be either all active, or have very close time-steps. However, it also occurs that neighboring particles are inactive, and this also produces either an artificial metal loss or gain.

To test the scenario of numerical gains/losses we analyse the evolution of an isolated disk galaxy under two different configurations. In the first case we impose that half of the gas in the disk has zero metallicity, whereas the other half has solar metallicity. This is an extreme scenario that maximizes the potential round off errors in the calculation of metal diffusion. In the second case, we impose that half of the gas in the disk that is randomly selected has zero metallicity, whereas the other half has solar metallicity.

\begin{figure*} 
\begin{center}
\includegraphics[angle=0,width=\textwidth]{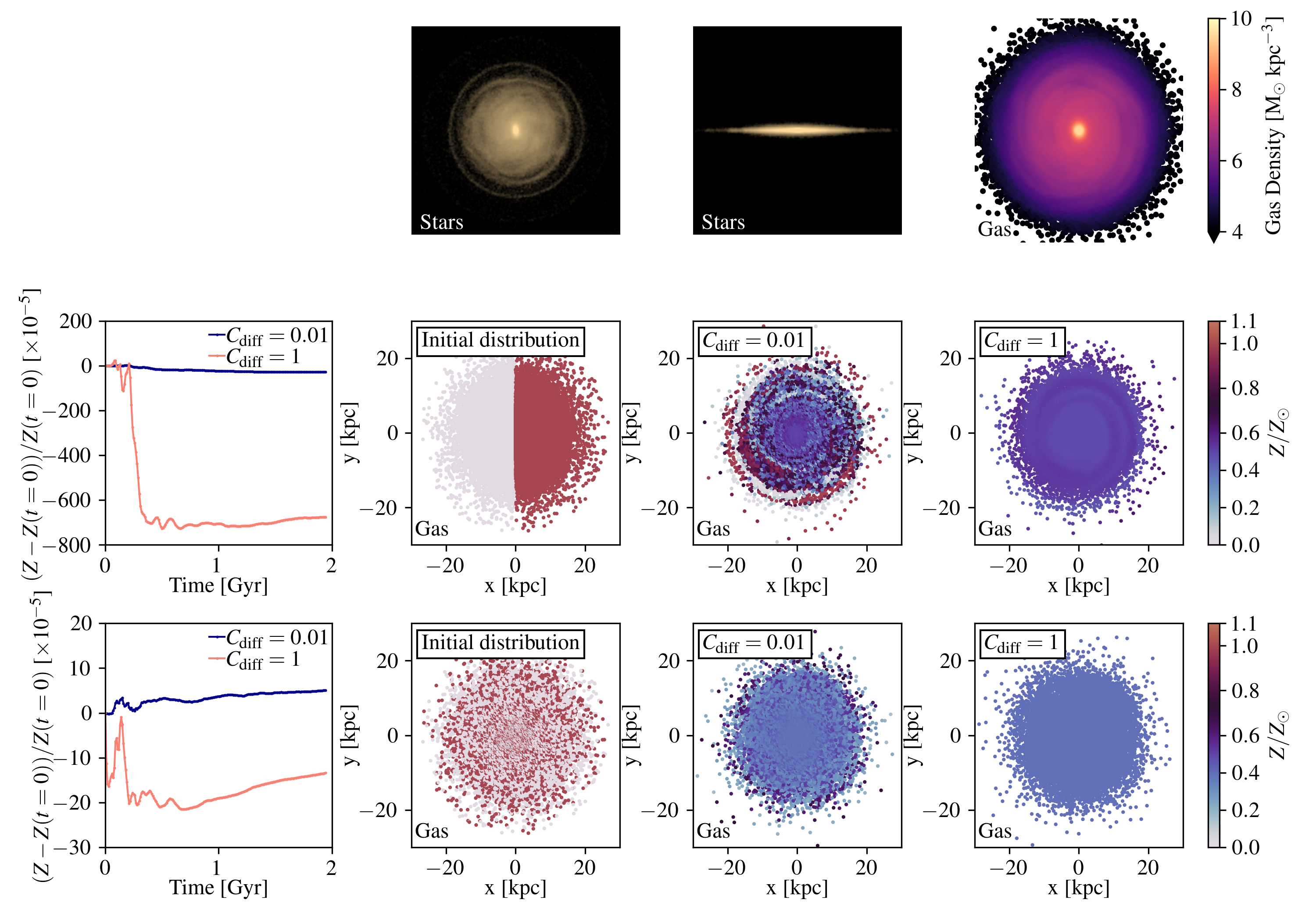}\\
\caption{Evolution of an isolated galaxy, where gas is set in hydrostatic equilibrium and it only evolves under gravity and metal diffusion. Top panels display the final distribution of stars (face and edge on) and gas (face on). Middle panels show the evolution of the metal content of the gas under a spatial symmetric metal configuration where half the gas had zero metallicity and the other half solar metallicity. The bottom panels show a similar scenario under the case that gas particles are randomly selected to have either zero or solar metallicity. In these two rows, the left column shows the evolution of the total metal content of the galaxy as a function of time. The second from the left shows the initial metal distribution, with gas particles coloured according to the colorbar on the right. The right and second from the right columns show the final metal distribution after 2 Gyr of evolution. These tests show that round off errors and particles under different time-steps can lead to numerical losses in the total metal content of up to 0.02\% for the default diffusion model.}
\label{Testsnap150}
\end{center}
\end{figure*}

The panels of Fig.~\ref{Testsnap150} show the isolated galaxy configuration. The top panels show the stellar disc (face on and edge on) and the density distribution of the gas (face on). For the initial conditions of the isolated galaxy we follow \citet{Nobels23}, the fiducial model with a particle mass of $10^5$ M$_{\odot}$ and a gravitational softening of 200 pc. We simulate the isolated galaxy formed by an exponential disk of gas and stars embedded in a dark matter halo modelled using an external gravitational potential. The dark halo follows a \citet{Hernquist90} profile with a virial mass of $1.37\times 10^{12}$ M$_{\odot}$. The disc of the galaxy has a mass of $3.8\times 10^{10}$ M$_{\odot}$ in stars and $1.6\times 10^{10}$ M$_{\odot}$ in gas. The gas is set to a temperature of $10^{4}$ K, and it is in vertical hydrostatic equilibrium, meaning that the scale height of the gas is set by the pressure and gravity of the gas and stars. We turn off all subgrid prescriptions for this test, meaning that cooling and star formation are disabled for the gas, as well as any form of feedback from the star particles. We only allow gas to evolve under gravity, hydrodynamics and metal diffusion.

The second panels from the left show the artificial initial configuration of the gas metallicity. The middle row shows that the gas in the negative x-direction has zero metallicity, whereas gas in the positive x-direction has solar metallicity. This isolated galaxy model is run for 2 Gyr under metal diffusion scheme with diffusion coefficient set to $C_{\rm{diff}}=0.01$ and $C_{\rm{diff}}=1$ (maximum diffusion). The right and second from the right panels in the middle row show the metal distribution of the gas after 2 Gyr. It can be seen that in the second from the right panel, after 2 Gyr of metal mixing, some gas particles reach a metallicity of $0.5$ Z$_{\odot}$, specially those in the central part of the disc, as evidenced by their color following the color bar on the right. Differently, particles in the outskirts either retained their full metal content or still had zero metallicity. This is due to the fact that the rate of metal mixing is dictated by the velocity shear of the gas, which in turn is a proxy for the gas turbulence. The higher the velocity shear, the larger the metal mixing rate. The right panel in the middle row shows that after 2 Gyr, almost all of the gas completely mixed and reached a metallicity of $\sim 0.5$ Z$_{\odot}$ when metal diffusion is set to maximum ($C_{\rm{diff}}=1$).

The left panel in the middle row shows the change in the total metallicity of the galaxy, $Z$, relative to its initial value, $Z(t=0)$ (in units of $10^{-5}$). It can be seen that under $C_{\rm{diff}}=0.01$, the total relative change in metallicity is smaller than $\Delta Z/Z<10^{-6}$. When $C_{\rm{diff}}$ is set to 1, the total relative change in metallicity is $\Delta Z/Z\sim -7\times 10^{-3}$. This means that under this extreme configuration, and when diffusion is maximum, the numerical losses in the galaxy's metal content, driven by round off errors, are of the order to $10^{-2}-10^{-3}$. 

The bottom panels of Fig.~\ref{Testsnap150} show a similar scenario, but in this case the initial metal abundance distribution is different. Gas is randomly selected to have either zero or solar metallicity. The bottom right and second from the right panels show that after 2 Gyr, almost all of the gas is fully mixed and reached a metallicity of $\sim 0.5$ Z$_{\odot}$. The bottom left panel shows that in this configuration, numerical losses reach $\Delta Z/Z\sim 5\times 10^{-5}$ and $\Delta Z/Z\sim -1.5\times 10^{-4}$ under $C_{\rm{diff}}=0.01$ and $C_{\rm{diff}}=1$, respectively. In this case, when $C_{\rm{diff}}=0.01$ there is an artificial creation of metals, and when $C_{\rm{diff}}=1$ some metals are lost.

From this test we conclude that for an extreme metallicity distribution, the default diffusion model (with $C_{\rm{diff}}=0.01$) can produce a creation or destruction of metals of up to 0.02\%, whereas the very high diffusion model (with $C_{\rm{diff}}=1$) can produce an artificial metal loss of up to 1\%. 

\bsp	
\label{lastpage}
\end{document}